\documentclass[prd,aps,showpacs,floats,letterpaper,floatfix,
groupedaddress,superscriptaddress
,nofootinbib
,eqsecnum,twocolumn
]{revtex4-1}

\usepackage{amssymb,amsmath,color}
\usepackage{graphicx}
\usepackage{dcolumn,epsfig}
\usepackage{hyperref}

\newcommand{\beq}{\begin{equation}}
\newcommand{\eeq}{\end{equation}}
\newcommand{\bes}{\begin{subequations}}
\newcommand{\ees}{\end{subequations}}
\newcommand{\bea}{\begin{eqnarray}}
\newcommand{\eea}{\end{eqnarray}}
\newcommand{\ba}{\begin{array}}
\newcommand{\ea}{\end{array}}
\newcommand{\beqn}{\begin{eqnarray*}}
\newcommand{\eeqn}{\end{eqnarray*}}

\newcommand{\f}[2]{\frac{#1}{#2}}

\newcommand{\LISA}{{\mbox{\tiny LISA}}}
\newcommand{\LIGO}{{\mbox{\tiny LIGO}}}
\newcommand{\ET}{{\mbox{\tiny ET}}}

\newcommand{\GW}{{\mbox{\tiny GW}}}
\newcommand{\ISCO}{{\mbox{\tiny ISCO}}}

\def\nn{\nonumber}

\def\be{\begin{equation}}
\def\ee{\end{equation}}
\def\beq{\begin{eqnarray}}
\def\eeq{\end{eqnarray}}

\def\IL{\relax{\rm I\kern-.18em L}}

\def\nn{\nonumber}
\def\f{\frac}

\begin{document}

\title{Post-Circular Expansion of Eccentric Binary Inspirals: \\
Fourier-Domain Waveforms in the Stationary Phase Approximation}

\author{Nicolas Yunes} \email{nyunes@princeton.edu} 
\affiliation{Department of Physics, Princeton University, Princeton, NJ 08544,
  USA}

\author{K. G. Arun} \email{arun@physics.wustl.edu}
\affiliation{McDonnell Center for the Space Sciences, Department of
Physics, Washington University, St.  Louis, Missouri 63130, USA}
\affiliation{GReCO, Institut d'Astrophysique de Paris, CNRS,\\ 
Universit\'e Pierre et Marie Curie, 98 bis Bd. Arago, 75014 Paris, France}
\affiliation{LAL, Universit\'e Paris Sud, IN2P3/CNRS, Orsay, France}

\author{Emanuele Berti} \email{berti@phy.olemiss.edu}
\affiliation{Department of Physics and Astronomy, The University of
Mississippi, University, MS 38677-1848, USA}
\affiliation{Theoretical Astrophysics 130-33, California Institute of
  Technology, Pasadena, CA 91125, USA}

\author{Clifford M.~Will} \email{cmw@wuphys.wustl.edu}
\affiliation{McDonnell Center for the Space Sciences, Department of
Physics, Washington University, St.  Louis, Missouri 63130, USA}
\affiliation{GReCO, Institut d'Astrophysique de Paris, CNRS,\\ 
Universit\'e Pierre et Marie Curie, 98 bis Bd. Arago, 75014 Paris, France}

\date{\today}

\begin{abstract}

We lay the foundations for the construction of analytic expressions for
Fourier-domain gravitational waveforms produced by eccentric, inspiraling
compact binaries in a post-circular or small-eccentricity approximation.  The
time-dependent, ``plus'' and ``cross'' polarizations are expanded in Bessel
functions, which are then self-consistently re-expanded in a power series
about zero initial eccentricity to eighth order.  The stationary phase
approximation is then employed to obtain explicit analytic expressions for the
Fourier transform of the post-circular expanded, time-domain signal. We
exemplify this framework by considering Newtonian-accurate waveforms, which in
the post-circular scheme give rise to higher harmonics of the orbital phase
and amplitude corrections both to the amplitude and the phase of the Fourier
domain waveform.  Such higher harmonics lead to an effective increase in the
inspiral mass reach of a detector as a function of the binary's eccentricity
$e_0$ at the time when the binary enters the detector sensitivity band. Using
the largest initial eccentricity allowed by our approximations ($e_0 < 0.4$),
the mass reach is found to be enhanced up to factors of approximately $5$
relative to that of circular binaries for Advanced LIGO, LISA, and the
proposed Einstein Telescope at a signal-to-noise ratio of ten.  A
post-Newtonian generalization of the post-circular scheme is also discussed,
which holds the promise to provide ``ready-to-use'' Fourier-domain waveforms
for data analysis of eccentric inspirals.

\end{abstract}
\maketitle

\section{Introduction}

The detection and characterization of gravitational waves (GWs) hold the
promise to reveal previously unattainable, yet very valuable astrophysical
information (see \cite{SathyaSchutzLivRev09} for a recent review).
Ground-based detectors, such as the Laser Interferometer Gravitational Wave
Observatory (LIGO)~\cite{LIGO_web}, VIRGO~\cite{VIRGO_web},
GEO~\cite{GEO_web}, and TAMA~\cite{TAMA_web}, have started acquiring data at
or near design sensitivity.  The space-borne Laser Interferometer Space
Antenna (LISA)~\cite{LISA_web} may be launched within the next decade, while
future Earth-based third-generation detectors, such as the proposed Einstein
Telescope (ET)~\cite{ET}, are currently being planned. These detectors are
expected to observe several different astrophysical GW sources, one of the
most promising of which are compact binary inspirals.

Black hole (BH) binaries are considered one of the main targets for GW
detection and their evolution can be roughly divided into an inspiral phase and a merger
plus ringdown phase. Peters and Mathews showed that eccentric inspirals
circularize via GW emission~\cite{Peters:1963ux,Peters:1964zz}, and thus, it
was traditionally thought that eccentricity would not play a major role in GW
detection and analysis. One can show,  for example, that to leading order
$e/e_{0} \sim (f/f_{0})^{-19/18}$~\cite{Mora:2003wt}, which implies that if a binary
enters the sensitivity band of a ground-based interferometer at $20$~Hz
with an initial eccentricity of $0.1$, its eccentricity is reduced to $0.01$
before the system reaches $200$~Hz.  Circularization via GW emission, however,
is not absolute, since systems that enter a detector's sensitivity band with
large enough eccentricity can retain some residual eccentricity before they
merge or exit the band. For example, if a binary enters the sensitivity band
at $20$~Hz with an initial eccentricity of $0.4$, its eccentricity remains
significant while in band, and is reduced to $0.01$ only by the time the 
frequency reaches $10^{3}$~Hz.

Astrophysical scenarios have been proposed that predict that binary inspiral
signals could enter the sensitivity band of GW detectors with non-negligible
eccentricity. More precisely, Earth-based detectors are expected to be sensitive to stellar
mass BH/BH binaries, which might fail to completely circularize before merger,
leading to potentially large eccentricities in the detector
band~\cite{Wen:2002km,OKL08}. Wen~\cite{Wen:2002km} studied the evolution of
the inner binary of triple BH systems in globular clusters. She found that
approximately $30 \%$ of these binaries could merge via the Kozai mechanism
with eccentricities $\geq 0.1$ by the time they enter the LIGO band at
frequency $\simeq 10$~Hz.  O'Leary {\it et al.}~\cite{OKL08} studied the density
cusp of stellar mass BHs that forms around supermassive BHs in galactic nuclei
because of mass segregation. They found that, in such dense environments,
hyperbolic BH-BH encounters can lead to the formation of bound systems, and
that most of these binaries ($\simeq 90 \%$) would have eccentricity $\geq 0.9$
when they enter the LIGO band.  Several classes of compact binary sources for
LISA are also predicted to be eccentric when they enter the detector's band.
In order to improve readability of this paper, we have relegated a brief
review of these astrophysical scenarios to Appendix~\ref{app:LISA}.

For binaries with non-negligible eccentricities, a dedicated search using
matched filtering techniques and eccentric orbit templates would be necessary
for detection and extraction of astrophysical information. Until now, however,
closed-form analytic expressions for such templates have been lacking, with
most studies concentrating on the circular case.  The modeling of eccentric
orbits is much more difficult, since it requires knowledge not only of the
orbital phase and frequency, but also of frequencies associated with the
higher harmonics of the eccentric motion, as well as of frequencies associated
with PN precession effects, such as pericenter precession.

Peters and Mathews' seminal work~\cite{PM63}, and Peters' follow-up
calculation of the angular momentum flux and evolution of orbital elements at
Newtonian order~\cite{Pe64}, laid the foundations for the calculation of
energy and angular momentum fluxes at higher PN
orders~\cite{WagW76,BS89,BS93,RS97,GI97,ABIQ07tail,ABIQ07,ABIS09}, the latter
mostly focusing on circular orbits. Analytic GW templates for circular
binaries are now available at 3.5PN order in the phase
\cite{BDIWW95,BFIJ02,BDEI04} and 3PN order in the amplitude
\cite{BIWW96,ABIQ04,KBI07,BFIS08}, and their associated Fourier transforms
have been computed in the stationary phase approximation.  Recently
Ref.~\cite{DGI04} provided a method to construct high accuracy templates for
elliptical binaries in the {\it time domain} by explicitly computing the
post-adiabatic short period contributions at 2.5PN order, to be added to the
post-Newtonian expressions for the GW polarizations. A 3.5PN generalization of
these templates was discussed in Ref.~\cite{KG06}.

Relatively few investigations of data analysis issues for eccentric
inspirals have been performed. Martel and Poisson ~\cite{Martel:1999tm} quantified the accuracy to
which circular orbit templates could capture signals from eccentric binaries
in the LIGO band, finding that the signal-to-noise ratio (SNR) loss is
significant for eccentricities above $0.1$.  Seto~\cite{Seto:2001pg} studied
parameter estimation in the context of eccentric galactic neutron star (NS)
binaries, and Benacquista~\cite{Benacquista:2001wa,Benacquista:2002kf} carried
out the first statistical investigation of the harmonic structure of eccentric
binary waveforms and their relevance for LISA GW detection.  The analysis by
Martel and Poisson was recently revisited~\cite{Tessmer:2007jg,Cokelaer:2009hj},
emphasizing the need for eccentric binary templates. All these investigations, 
however, concentrated on either time-domain waveforms or numerical Fourier transforms of
such waveforms, which might not be desirable for data analysis purposes.

The aim of this paper is to lay the foundations of a post-circular
approximation that allows for the construction of analytic, ``ready-to-use''
Fourier-domain waveforms for eccentric binary inspirals. This approximation
consists of expanding time-domain gravitational waveforms in the
eccentricity parameter $e_0$, which is defined to be the eccentricity when 
the GW signal enters the detector sensitivity band. The resulting expression
in then Fourier transformed in the stationary-phase approximation. This scheme is an extension of the
program initiated by Krolak, Kokkotas and Sch\"afer several years
ago~\cite{Krolak:1995md} and it is intended to supplement the PN
approximation, thus yielding a double or {\emph{bivariate}} expansion in both
the velocity of the binary members and the initial eccentricity.  Although the
PN scheme does not require the post-circular approximation for the
construction of numerical Fourier-domain templates, closed-form analytic
expressions for these templates can only be obtained through the incorporation
of the post-circular approximation.

The usefulness of ``ready-to-use,'' analytic, frequency-domain waveforms is
two-fold.  Analytic expressions allow us to study the structure of the
eccentricity induced corrections to the Fourier transform of the signal; in
turn, this structure allows us to explain features in the SNR that would
otherwise be hidden by numerics. Secondly, analytic expressions allow for fast
implementations of dedicated matched-filtering searches in a data analysis
algorithm and allow us to sidestep fast Fourier transforms, which would drain
numerical resources from Fourier-domain data analysis pipelines.

We exemplify the post-circular approximation by considering Newtonian 
expressions for the two GW polarization states of elliptic binaries. 
The cosines and sines of the GW phase are expanded in a truncated Bessel
series, whose argument is proportional to the eccentricity parameter. We find
that the first $9$ terms in the sum suffice to approximate the phase to better
than $0.1 \%$ for eccentricities smaller than $0.4$ (see Sec.~\ref{Sec:kepler}
below).  The Bessel series is then re-expanded to eighth order in $e \ll 1$,
which we find sufficient to capture the essential features of orbital dynamics
for such small eccentricities.

The structure of the time-domain gravitational waveform for eccentric 
inspirals takes the form 
\be
h_{+,\times} \sim F^{2/3} \sum_{\ell=1}^{10} 
\left[ C_{+,\times}^{(\ell)} \cos{(\ell l)} + S_{+,\times}^{(\ell)} \sin{(\ell l)} \right],
\label{struct}
\ee
where $F$ is the orbital frequency, $l$ is the mean anomaly and $C_{+,\times}$
and $S_{+,\times}$ are eighth-order power series in the eccentricity [see
  Eqs.~(\ref{CS1})-(\ref{CS1end}) and Eqs.~(\ref{C+})-(\ref{Sx})]. The
eccentricity is itself a function of the orbital frequency, which we invert in
the limit $e \ll 1$ and insert into the time-domain waveforms to obtain
explicit expressions whose only independent variables are the orbital
frequency and the mean anomaly.  The prescription described above to compute
time-domain waveforms is a completion of the analytic work of Moreno-Garrido,
Buitrago and Mediavilla~\cite{1994MNRAS.266...16M,1995MNRAS.274..115M},
improved through the resummation methods of Pierro and
Pinto~\cite{Pierro:2000ej,Pierro:2002wd} and the more general waveform
expressions of Martel and Poisson~\cite{Martel:1999tm}.

Once closed-form, analytic expressions for the time-domain waveforms are
obtained, we compute their Fourier transform in the stationary phase
approximation (SPA). This approximation derives from the asymptotic method of
integration by steepest descent, which allows one to systematically include
higher harmonics in the frequency-domain waveforms. This higher-harmonic
structure is found to fit perfectly in the formalism of
Refs.~\cite{ChrisAnand06,Arun:2007qv}, which was developed to account for
higher harmonics due to PN amplitude corrections in circular-orbit binary
waveforms.

The Fourier transform of the response function is then found to take the form
\be
\tilde{h} \sim \tilde{\cal{A}} \; f^{-7/6} \sum_{\ell=1}^{10}  \left({\ell\over2}\right)^{2/3} \xi_{\ell} \; e^{-i  \psi_{\ell}},
\ee
where $\tilde{\cal{A}}$ is an overall amplitude that depends on the system
parameters (such as the masses of the binary members), while $f$ is the
dominant (quadrupole) GW frequency. The amplitudes $\xi_{\ell}$ and the phases
$\psi_{\ell}$ are small-eccentricity expansions [see
  Eqs.~\eqref{final-new-Kostas} and~\eqref{xis}]. The $\xi_{\ell}$'s depend on
the antenna pattern functions $F_{+,\times}$ \cite{Th300}, the initial
eccentricity $e_{0}$ and the GW frequency.  The expansion of the phase reads
\be
\psi_{\ell} \sim - \frac{3}{128} f^{-5/3} \left(\frac{\ell}{2}\right)^{8/3} \left[ 1 - \frac{2355}{1462} e_{0}^{2} \left(\frac{f}{f_{0}}\right)^{-19/9} + \ldots \right],
\ee
where $f_{0}$ is the frequency at which the eccentricity equals $e_{0}$, which
we choose to coincide with the low-frequency cut-off of
the detector sensitivity band. This is in agreement with the Newtonian limit
of Eq.~(A9) of \cite{Krolak:1995md} up to ${\cal O}(e_0^2)$.

\begin{figure}
\includegraphics[width=8cm,clip=true]{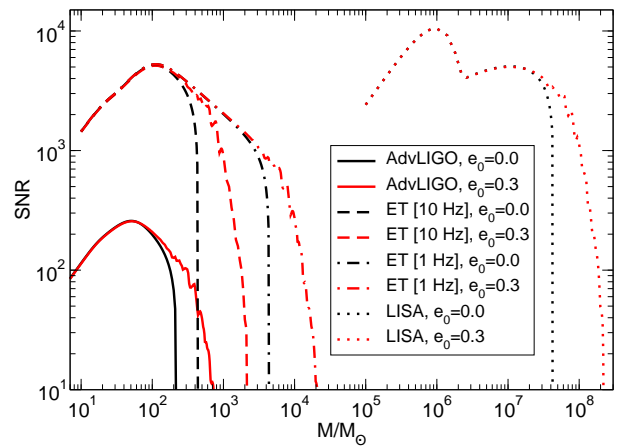}
\caption{\label{AL-ET-LISA} SNR for an equal-mass binary at optimal
  orientation as a function of total mass for circular binaries and elliptic
  binaries with initial eccentricity of $0.3$. The assumed low-frequency
  cut-offs for AdvLIGO, ET and LISA are $20$~Hz, either $1$ or $10$~Hz, and
  $10^{-4}$~Hz, respectively.  The sources are at $100$ Mpc for AdvLIGO and ET
  and at $3$ Gpc for LISA. The initial eccentricity corresponds to that at the
  low-frequency cut-off.}
\end{figure}

One of the benefits of obtaining closed-form, analytic expressions for the
Fourier transform of the waveforms is that its harmonic structure and its
eccentricity-induced amplitude corrections become explicit. We find that these
higher-harmonic eccentricity corrections increase the SNR for large total
masses, in analogy to what was found for PN amplitude-corrected circular orbit
waveforms in~\cite{ChrisAnand06,Arun:2007qv,TriasSintes07}.
Figure~\ref{AL-ET-LISA} compares the optimal SNR of equal-mass binaries with
eccentricity $e_{0} = 0$ and $e_{0}= 0.3$ at the initial frequency of the
sensitivity band of Advanced LIGO (AdvLIGO), ET and LISA ($20$~Hz, $1$~Hz or $10$~Hz, and 
$10^{-4}$~Hz respectively). The source is located at distances
of $100$ Mpc for AdvLIGO and ET, and $3$ Gpc for LISA. By ``optimal'' we mean
the SNR measured by an observer located in a direction perpendicular to the
orbital plane (more precisely, we set $\iota=\beta=\theta_S=\phi_s=\psi_S=0$
in the notation of Section~\ref{Sec:model}).  This SNR increase is rather
generic, irrespective of the location of the source in the sky.

\begin{figure}
\includegraphics[width=8cm,clip=true]{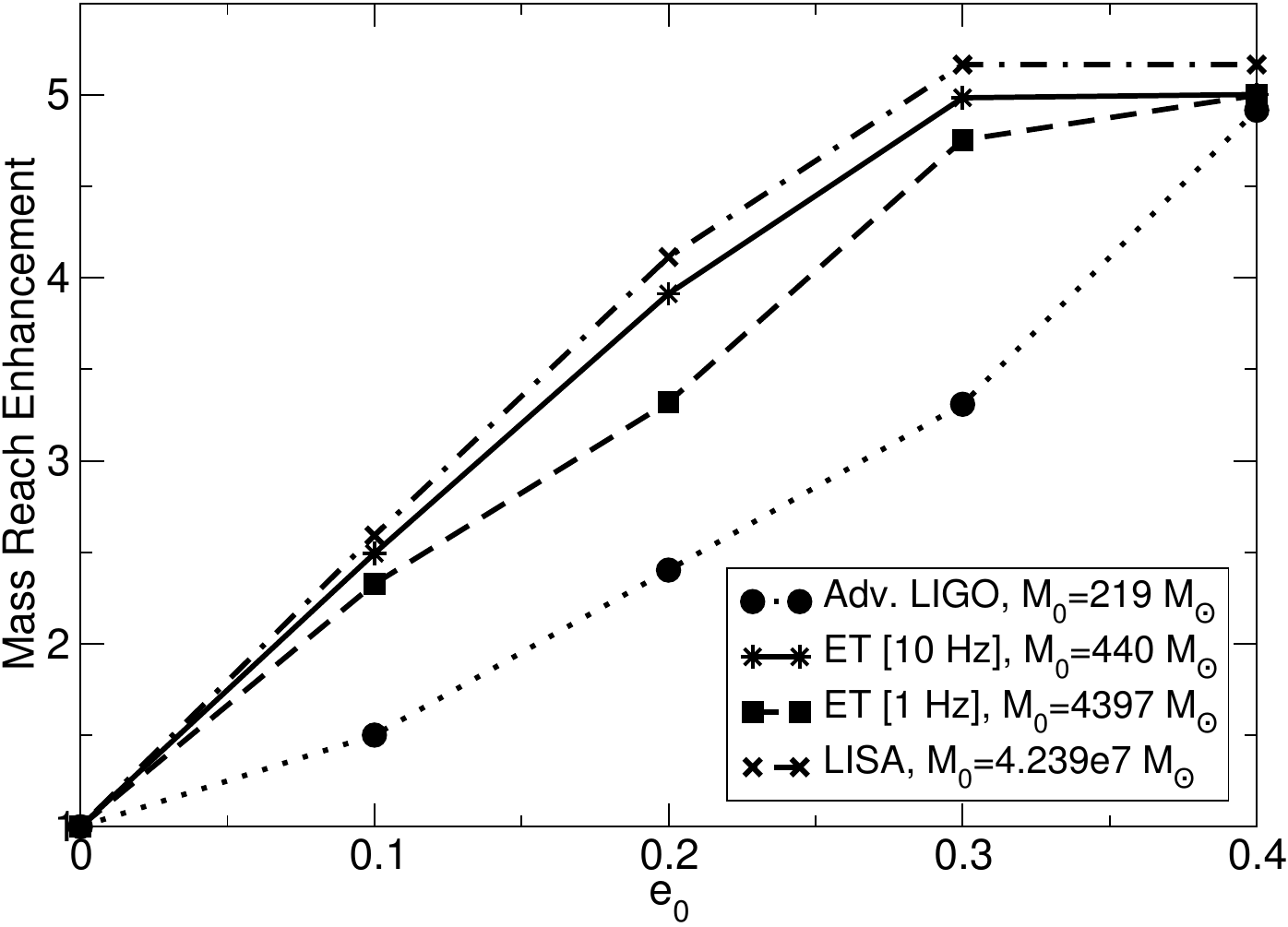}
\caption{\label{Eplot} Normalized mass reach enhancement, as a
  function of initial eccentricity. The normalization is given by the value of
  the mass reach for circular binaries, namely $M_{0}=219 M_{\odot}$,
  $M_{0}=440 M_{\odot} (4397 M_{\odot})$ and $M_{0}=4.239 \times 10^{7}
  M_{\odot}$ for AdvLIGO, ET and LISA, respectively.}
\end{figure}

The inclusion of eccentricity in the waveforms leads to an increase in the
mass reach as compared to circular waveforms. This is shown in
Fig.~\ref{Eplot}, where we plot the mass reach enhancement $M(e_{0})/M_{0}$,
where $M_{0}=M(0)$. The mass reach $M(e_{0})$ is here defined as the mass
corresponding to an optimal SNR of ten, roughly corresponding to the largest
mass visible to the detector.
The mass reach usually increases with $e_0$, up to factors of order five for
binaries with $e_{0} \simeq 0.4$. This result should still hold when PN
corrections are included, since we expect these corrections to increase the
mass reach. We conclude then that LISA could potentially observe moderately
eccentric binaries with total masses on the order of $10^{8} M_{\odot}$. These
results are in agreement with preliminary results from numerical relativity simulations
of merging eccentric binaries ~\cite{birjoo} (see also
\cite{Berentzen:2008yw}).

An increase in mass reach in turn implies that, for a system of fixed total
mass, the distance to which the system can be observed also increases with
eccentricity. For example, IMBH mergers of total mass $\simeq 200M_\odot$ with
orbital eccentricity of $e_0\simeq 0.3$ would be observable by AdvLIGO up to
approximately $1.26$~Gpc ($z\simeq 0.26$) with an SNR of 10, while SMBHs of
total mass $\simeq 4 \times 10^{7} M_{\odot}$ and initial eccentricity $e_{0}
\simeq 0.3$ would be visible by LISA up to approximately $100$~Gpc ($z\simeq
10$) with an SNR of $100$.  Such an increase in distance corresponds to an increase in accessible
volume of up to a factor of approximately one hundred for systems with $e_{0}
\simeq 0.4$.

The post-circular approximation also allows us to determine how many harmonics
are needed to reproduce the Fourier transform of the eccentric signal to some
accuracy. For a system with $e_{0} = 0.01$, we find that keeping up to the
second or third harmonic suffices to reproduce the SNR of a signal that
includes ten harmonics to ${\cal{O}}(1)$ and ${\cal{O}}(10^{-1})$ respectively
in the entire mass range.  For a system with $e_{0} = 0.1$, however, we find
that a comparable accuracy in SNR requires including up to the fourth and
fifth harmonic, respectively.  Such an analysis allows us to conclude that the
inclusion of up to the fourth harmonic suffices for SNR calculations when
$e_{0} \leq 0.1$, while for systems with $0.1 < e_{0} \leq 0.3$, one must
really include eight harmonics or more.

Although the waveforms we consider in our post-circular scheme are not
accurate enough for a rigorous data analysis study, as we have ignored PN
effects, they do provide insight as to the effect of eccentricity in detection
and parameter estimation. As for detection, the SNRs presented here may well
be smaller than SNRs for PN-corrected eccentric binary inspirals, since the
addition of harmonics to the amplitudes generally increases the power in the
signal.
As for parameter estimation, the eccentricity corrections to the phase of the
Fourier transform have a different frequency dependence relative to PN
corrections to the phase in the circular case. This suggests that the initial
eccentricity might be weakly correlated to other intrinsic parameters. A more
detailed study is necessary to verify this conjecture, and it will be a topic
for future work. In Section \ref{Sec:PN} we outline a possible extension of
the formalism to higher PN orders. When this extension is achieved,
ready-to-use Fourier domain gravitational waveforms could be employed in GW
searches and parameter estimation.

The remainder of this paper deals with the details of the calculations and
results presented above. It is organized as follows. Section~\ref{Sec:kepler}
presents the basics of the Kepler problem and establishes the notation used in
this paper. Section~\ref{Sec:model} discusses how to model GWs from eccentric
binary inspirals in the time domain, while Sec.~\ref{Sec:SPA} describes its
frequency-domain representation. Section~\ref{Sec:SNR} presents the SNR
calculation, while Section~\ref{Sec:PN} discusses PN
corrections. Section~\ref{Sec:conclusions} concludes and points to future
research.

Technical details are discussed in the Appendices. Appendix \ref{app:LISA}
reviews astrophysical scenarios that could produce eccentric binaries in the
LISA band. Appendices \ref{app:ccoeffs} and \ref{app:xik} list some lengthy
coefficients appearing in our analytic calculations.  Appendix \ref{app:ISCO}
discusses the effect of possible eccentricity-induced modifications to the
innermost-stable circular orbit (ISCO), concluding that they are negligible in
our context. Finally, Appendix \ref{Fyr-app} shows how to compute the orbital
frequency at some given time before merger for eccentric binaries.

In this work we follow the conventions of Misner, Thorne and Wheeler
\cite{Misner:1974qy}: the metric has signature $(-,+,+,+)$; spacetime indices
are labeled with Greek letters, while spatial indices are labeled with Latin
letters; unless otherwise specified, we use {\emph{geometrical units}}, where
$G=c=1$, $G$ stands for Newton's gravitational constant and $c$ for the speed
of light.

\section{The basics of the Kepler problem} 
\label{Sec:kepler}

In this section, we review some of the basic concepts related to the Kepler
problem in Newtonian mechanics, as they are relevant to this paper.  We
present here only a minimal description of this problem and refer the reader
to~\cite{Goldstein} for a more detailed account. We also establish the
notation we shall employ in the remainder of this paper.

Consider a system of two point particles in an eccentric orbit.  In the
Newtonian Keplerian representation, the Newtonian orbital trajectories are
given by
\beq
\label{radius-eq}
r &=& a \left(1 - e \cos{u}\right),
\\
\label{radius-eq2}
N \left(t - t_0\right) &=& l = u - e \sin{u},
\\ 
\label{radius-eq3}
\phi - \phi_0 &=& v\equiv 2\arctan
\left[\left(\f{1+e}{1-e}\right)^{1/2}\tan\f{u}{2}
\right],
\eeq
where the notation, following~\cite{Gopakumar:2001dy}, is as follows: $\phi$
is the orbital phase; $\vec{r}$ is the relative separation vector between the
compact objects, namely $\vec{r}=r(\cos\phi\,,\sin\phi\,,0)$; $a$ is the
semi-major axis of the ellipse; $e$ is the eccentricity parameter; $u$ is the
eccentric anomaly; $l$ is the mean anomaly; $v$ is the true anomaly, and $N$
is the mean motion. The quantities $t_0$ and $\phi_0$ are some initial time
and initial orbital phase that arise as constants of integration.  Since the
energy and angular momentum fluxes depend on $a$ and $e$, and together cause
the latter to vary with time, it can be shown that $a$ and $e$ co-evolve
according to~\cite{Peters:1963ux,Peters:1964zz}
\be
\label{c0}
a(e)=\left(\f{M}{4 \pi^{2} F^2}\right)^{1/3}=c_0 \sigma(e)\,,
\ee
where $M=m_{1}+m_{2}$ is the total mass. 
The quantity $F$ is the Keplerian mean orbital frequency, which
can be associated with an instantaneous mean orbital frequency 
whose evolution is discuss further in Sec.~\ref{ecc-case-FD}.
The quantity $c_0$ is a constant defined by $F(e_0) = F_0$ and the function $\sigma(e)$ is given by
\be
\sigma(e)=\f{e^{12/19}}{(1-e^2)}\left[1+\f{121}{304}e^2\right]^{870/2299}\,.
\ee
The quantity $e_{0}$ is henceforth always defined to be the eccentricity when the GW signal 
enters the detector sensitivity band. For AdvLIGO, ET and LISA, this corresponds
to the initial eccentricity at $20$~Hz, $1$~Hz or $10$~Hz, and $10^{-4}$~Hz respectively. 

Gravitational waveforms for binary inspirals depend on trigonometric functions
of the orbital phase, but for eccentric inspirals this phase is a complicated
function of the orbital frequency. In the circular orbit limit ($e \to 0$),
the orbital phase satisfies $\phi = 2 \pi F(t) (t - t_0)$, where $F$ is the
Keplerian orbital frequency (one half the dominant, quadrupole GW frequency). 
For eccentric inspirals, however, the phase is related to the arctangent of
the eccentric anomaly, which is then related in a transcendental way to the
mean motion, and thus, to the frequency $N = 2 \pi F$. We must then find a way
to express the orbital phase as a function of the mean anomaly $l$.

Let us re-express the cosine and sine of the orbital phase in terms of the
mean anomaly, through the well-known Keplerian relations \cite{Pierro:2000ej}
\beq
\frac{r}{a} \cos{\phi} &=& \cos{u} - e,
\\
\frac{r}{a} \sin{\phi} &=& \left(1 - e^2\right)^{1/2} \sin{u}.
\eeq
Equation~(\ref{radius-eq}) allows us to rewrite these relations as
\beq
\label{phase-eqns:1}
\cos{\phi} &=& \frac{\cos{u} - e}{1 - e \cos{u}},
\\
\label{phase-eqns:2}
\sin{\phi} &=& \left(1 - e^2\right)^{1/2} \frac{\sin{u}}{1 - e\cos{u}}.
\eeq
Moreover, from the Fourier analysis of the Kepler problem, one can expand
trigonometric functions of the eccentric anomaly as series of Bessel functions
of the first kind, $J_{k}$. One then finds that~\cite{Pierro:2000ej}
\beq
\frac{\sin{u}}{1 - e\cos{u}} &=& 2 \sum_{k=1}^{\infty} J'_k(ke)
\sin{kl},
\label{u-eqs:1}
\\
\frac{\cos{u}}{1 - e\cos{u}} &=& \frac{2}{e} \sum_{k=1}^{\infty} J_k(ke)
\cos{kl},
\label{u-eqs:2}
\eeq
where primes stand for derivatives with respect to the argument, and
\be
J_k(y) \equiv \sum_{m=0}^{\infty} \frac{(-1)^m}{m! \Gamma(m+k+1)} \left(
  \frac{y}{2} \right)^{2 m + k},
\ee
with $\Gamma$ the Gamma function (see e.g.~\cite{AS}).

With these relations, we can now express the cosine and sine of the orbital
phase as a function of the mean anomaly. Inserting Eqs.~\eqref{u-eqs:1}
and~\eqref{u-eqs:2} into Eqs.~\eqref{phase-eqns:1} and~\eqref{phase-eqns:2} we
find that
\beq
\label{moreno1}
\cos{\phi} &=& -e + \frac{2}{e} \left(1 - e^2 \right)
\sum_{k=1}^{\infty} J_k(ke) \cos{k l},
\\
\label{moreno2}
\sin{\phi} &=& \left(1 - e^2 \right)^{1/2} \sum_{k=1}^{\infty}
\left[J_{k-1}(ke) - J_{k+1}(ke) \right] \sin{kl}.
\nonumber \\
\eeq
Equations~(\ref{moreno1}) and~(\ref{moreno2}) agree with the corresponding
expressions in the Appendix of Ref.~\cite{1994MNRAS.266...16M}. These
relations allow us to express the gravitational waveforms for eccentric
inspirals as explicit functions of the orbital frequency.

\begin{figure}
\includegraphics[width=8cm,clip=true]{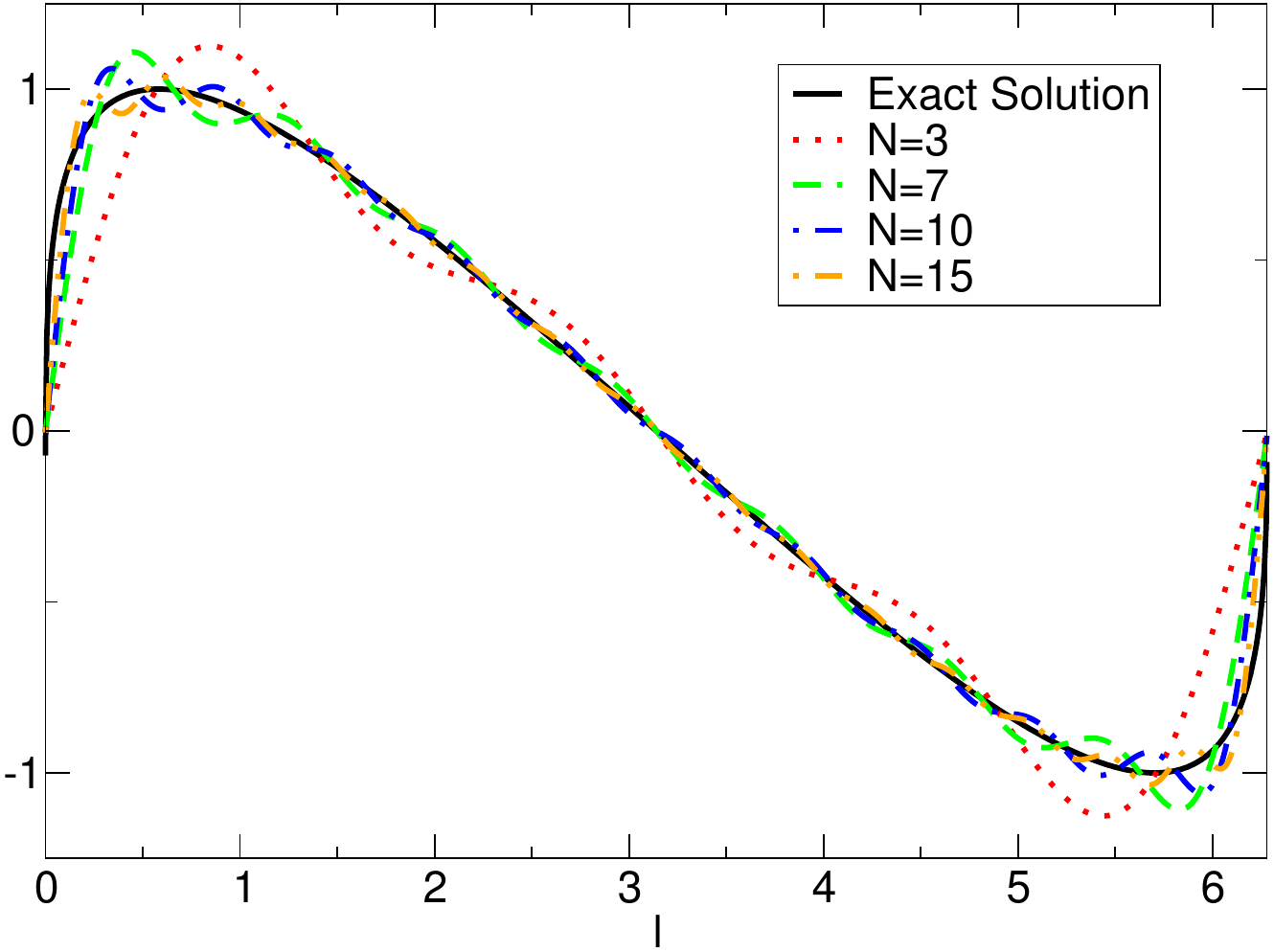}
\caption{\label{bessel} Plot of the sine of the phase calculated numerically
  (solid black) and expanded in Bessel functions for a system with $e_{0} =
  0.99$, following Eq.~\eqref{moreno2}, and keeping $3$ (dotted red), $7$
  (dashed green), $10$ (dot dashed blue) and $15$ (dot-dot-dashed orange)
  terms.}
\end{figure}

The number of terms we should keep in the Bessel function expansion depends on
the accuracy desired relative to the exact solution, as well as on the
magnitude of the eccentricity. Figure~\ref{bessel} plots the numerical
solution of Eq.~(\ref{phase-eqns:1}) for the sine of the orbital phase with an
eccentricity of $0.99$, together with the Bessel expansion of the solution
given in Eq.~(\ref{moreno2}), where we keep $3$ (dotted), $7$ (dashed), $10$
(dot-dashed) and $15$ (dot-dot-dashed) terms in the sum. Observe that even for
such large eccentricities we only need fewer than $10$ terms to reproduce the
exact solution quite accurately.

\begin{figure}
\includegraphics[width=8cm,clip=true]{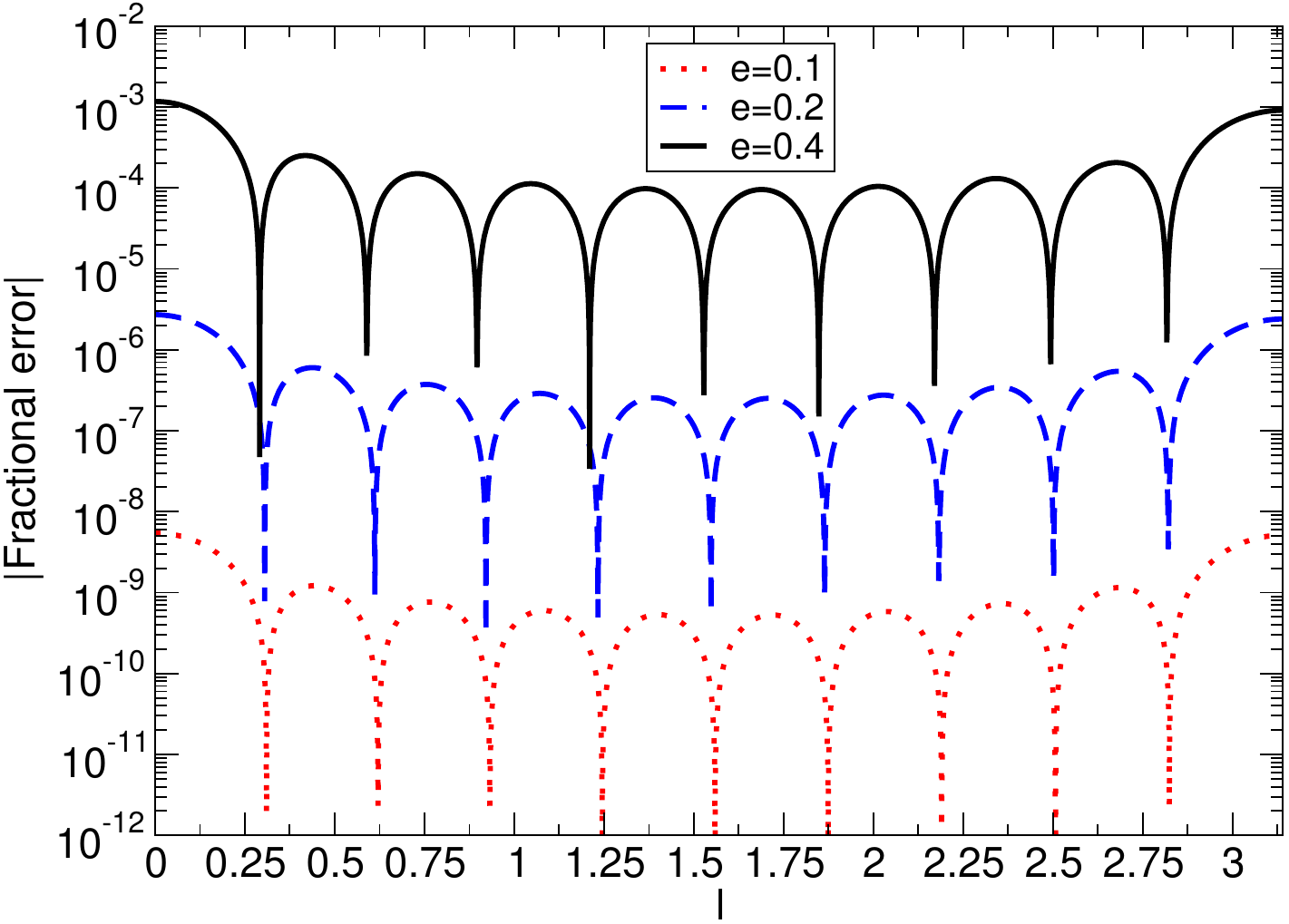}
\caption{\label{difference} Plot of the absolute value of the fractional
  relative difference between the sine of the eccentric anomaly, calculated
  numerically and expanded in Bessel functions with $9$ terms, for different
  eccentricities ($e=0.4$ solid black curve, $e=0.2$ blue dashed curve,
  $e=0.1$ red dotted curve). The results are here normalized by the exact
  numerical solution.}
\end{figure}

The number of terms needed to reproduce the exact solution does depend on the
eccentricity one is trying to model. As an example, consider solving
Eq.~(\ref{radius-eq2}) for the sine of the eccentric anomaly both numerically
and with Bessel functions. The latter solution is simply given
by~\cite{Pierro:2000ej}
\be
\sin{u} = \frac{2}{e} \sum_{k=1}^\infty J_k(k e) \frac{\sin{k l}}{k}.
\ee
In Fig.~\ref{difference} we plot the absolute value of the fractional relative
difference between the numerical solution and the Bessel-expanded solution,
keeping $9$ terms in the sum. We plot the absolute value of the difference,
normalized by the numerical solution for different eccentricities: $e=0.1$ in
dotted red, $e=0.2$ in dashed blue, and $e=0.4$ in solid black.
Observe that the error due to neglecting terms beyond the $9$th in the sum
amount to less than $0.1 \%$ in the worst case (corresponding to the highest
eccentricity $e=0.4$). For cases with smaller eccentricity, such as $e=0.2$
and $e=0.1$, the relative fractional error is always much smaller ($10^{-5}$
and $10^{-8}$ for the examples above, respectively).

The small-eccentricity assumption could be removed by working directly with
the full series in Eqs.~\eqref{moreno1} and~\eqref{moreno2}, or by resumming
it. In fact, Pierro and Pinto~\cite{Pierro:2000ej} have shown how to sum the
infinite Bessel series to measure the so-called ``total harmonic distortion,''
which is loosely related to Apostolatos' fitting
factor~\cite{Apostolatos:1995pj}.  We shall not, however, work with such
resummations here, since we shall be interested in binaries with
eccentricities $e < 0.4$. In view of this, we shall truncate all expressions
at order ten in the Bessel expansions.

\section{Post-circular Expansion of Time-Domain Eccentric Inspiral Waveforms} 
\label{Sec:model}

In this section we describe how to model gravitational radiation from
eccentric inspiraling binaries in the time domain. We shall employ the
quadrupole formalism, similar to that presented by Moreno-Garrido, Buitrago
and Mediavilla~\cite{1994MNRAS.266...16M,1995MNRAS.274..115M}, but improved
through the techniques introduced by Pierro and
Pinto~\cite{Pierro:2000ej,Pierro:2002wd}.

The starting point is the expression for plus- and cross-polarized
gravitational waveforms, $h_+$ and $h_\times$.  We place the GW detector at
luminosity distance $D_L$ from the source, in a direction characterized by the
polar angles $\iota$ and $\beta$, defined as those subtended by the local
Cartesian reference frame of the source and the line of sight
vector~\cite{Martel:1999tm}.  Following Martel and
Poisson~\cite{Martel:1999tm}, we rewrite the expressions of
Wahlquist~\cite{Wahlquist:1987rx} as follows
\beq\label{MP}
h_+&=&-\f{\mu}{pD_L}\left\{
  \left[2\cos(2\phi-2\beta)+\f{5e}{2}\cos(\phi-2\beta) 
\right. \right. 
\nonumber \\
&+& \left. \left. 
\f{e}{2}\cos(3\phi-2\beta)+e^2\cos(2\beta) \right](1+\cos^2\iota) 
\right. 
\nonumber \\
&+& \left. 
\left[e\cos\phi+e^2\right]\sin^2\iota \right\}\,,
\nn \\
h_\times&=&-\f{\mu}{pD_L}\left[
  4\sin(2\phi-2\beta)+5e\sin(\phi-2\beta) 
\right. 
\nonumber \\
&+& \left. 
e\sin(3\phi-2\beta) -2e^2\sin(2\beta) \right]\cos\iota\,,
\eeq
where $\mu=m_{1} m_{2}/M$ is the reduced mass, $D_{L}$ is the luminosity
distance, $\phi$ is the same orbital phase presented in Eq.~\eqref{radius-eq3}
and $p$ is the semi-latus rectum, which is related to the orbital frequency
via Kepler's second law
\be
\frac{1}{F} = 2 \pi M^{-1/2} \left( \frac{p}{1 - e^2} \right)^{3/2}.
\ee

GWs are clearly dominated, at least for small eccentricities, by components
oscillating at once, twice and three times the orbital frequency.
Equation~(\ref{MP}) is valid in the quadrupole approximation, which means that
higher multipoles (the octupole, hexadecapole and higher) have been neglected.
These multipoles are proportional to terms of ${\cal{O}}(\dot{r}/c)$ and
higher, which implies that Eq.~\eqref{MP} is a good approximation for slow
velocities and weak gravity.

The harmonic structure discussed above can be seen more clearly if we
re-express Eq.~\eqref{MP}, by using trigonometric identities, as:
\begin{widetext}
\beq
\label{h+-harm}
h_+ &=& \frac{\cal{A}}{1 - e^2} \left\{ \cos{\phi} \left[ e s_i^2
      + \frac{5 e}{2} c_{2 \beta} \left(1 + c_i^2\right)\right] + \sin{\phi}
    \left[ \frac{5 e}{2} s_{2 \beta} \left(1 +
      c_i^2\right) \right] + \cos{2 \phi} \left[ 2
    c_{2 \beta} \left( 1 + c_i^2 \right) \right] 
\right.
\\
&+& \left.  \sin{2 \phi} \left[
    2 s_{2 \beta} \left(1 + c_i^2\right) \right] +
  \cos{3 \phi} \left[ \frac{e}{2} c_{2 \beta}
    \left(1 + c_i^2\right) \right] + \sin{3 \phi} \left[ \frac{e}{2}  
    s_{2 \beta} \left(1 + c_i^2\right)\right]
\nonumber 
+ e^2 s_i^2 + e^2  \left(1 + c_i^2\right) c_{2
      \beta} \right\},
\\
\label{hc-harm}
h_{\times} &=& \frac{\cal{A}}{1 - e^2} \left\{ \cos{\phi} \left[ - 5 e
    s_{2 \beta} c_i \right] + \sin{\phi} \left[ 5 e
    c_{2 \beta} c_i \right] + \cos{2 \phi} \left[- 4
    s_{2 \beta} c_i \right] + \sin{2 \phi} \left[ 4
    c_{2 \beta} c_i \right]
\right. 
\\
&+& \left.   
 \cos{3 \phi} \left[ - e
    s_{2 \beta} c_i \right] + \sin{3 \phi} \left[ e
    c_{2 \beta} c_i \right] - 2 e^2 s_{2 \beta} c_i \right\},
\nonumber 
\eeq
\end{widetext}
where we defined $c_i \equiv \cos{\iota}$, $s_i \equiv \sin{\iota}$, $c_{2
  \beta} \equiv \cos{2 \beta}$ and $s_{2 \beta} \equiv \sin{ 2 \beta}$, and we
introduced the amplitude
\be\label{amplitude}
{\cal{A}} \equiv -\frac{{\cal M}}{D_L} \left( 2 \pi {\cal M} F\right)^{2/3},
\ee
with the chirp mass given by ${\cal{M}} \equiv \mu^{3/5} M^{2/5}$.  In the
limit $e \ll 1$ the dominant term is the second harmonic, followed by the
first and third harmonics, while the constant term contributes only to higher
order.

Explicit expressions for the waveforms as functions of time are needed to
construct their Fourier transform. We shall thus substitute the expansions of
the sines and cosines of the phase in terms of Bessel functions
[Eqs.~\eqref{moreno1} and~\eqref{moreno2}] into Eqs.~(\ref{h+-harm})
and~(\ref{hc-harm}).  The Bessel functions, however, are themselves
polynomials in the eccentricity, which we are also expanding about.  Three
different expansions are thus taking place:
\begin{itemize}

\item{\bf{PN and Multipole Expansion}}: Weak-field expansion of the metric in
  terms of mass and current multipole moments of the source distribution,
  which are then expanded in small-velocities.

\item {\bf{Bessel Expansion}}: Expansion of the orbital phase in Bessel
  coefficients.

\item {\bf{Eccentricity Expansion}}: expansion of Bessel coefficients in small
  eccentricities.

\end{itemize}

The multipolar expansion is a weak-field expansion to solutions to the
Einstein equations, where we shall here keep only the mass quadrupole. This
implies that our waveforms are accurate only to Newtonian order, where we
neglect terms of relative order ${\cal{O}}(\dot{r}/c)$. Such terms can be
accounted for through a PN analysis, as we shall discuss in Sec.~\ref{Sec:PN}.

The Bessel and eccentricity expansions are related, and one must ensure that
they are performed to a consistent order. Bessel functions of the first kind
behave as $J_{k}(ke) \sim e^{k}$ asymptotically for $e \ll 1$. The phases in
Eqs.~\eqref{moreno1} and~\eqref{moreno2}, however, scale as $\sin{\phi} \sim
\cos{\phi} \sim J_{k}(ke)/e \sim e^{k-1}$. Thus, an expansion of the waveforms
to ${\cal{O}}(e^{N})$ requires the phases in Eqs.~\eqref{moreno1}
and~\eqref{moreno2} to be summed up to $k_{\textrm{max}} = N+1$.  We shall
here work to $N=8$, which means that the Bessel sums in Eqs.~\eqref{moreno1}
and~\eqref{moreno2} must be performed up to $k_{\textrm{max}}=9$.

With these expansions, the waveforms can be written as a sum over harmonics. 
Using that the mean anomaly $l =2 \pi F (t - t_0)$, the waveforms become
\be
\label{expansion}
h_{+,\times} = 
{\cal{A}} \sum_{\ell=1}^{10} \left[C_{+,\times}^{(\ell)} \cos{\left(\ell l\right)} 
+ S_{+,\times}^{(\ell)} \sin{\left(\ell l \right)} \right],   
\ee
where the $\ell=1$ coefficients are 
\beq
\label{CS1}
C_+^{(1)} &=& s_i^2 \left(e - \frac{1}{8} e^3 + \frac{1}{192} e^5 -
  \frac{1}{9216} e^7 \right) 
  \\ \nonumber 
  &+& \left(1 + c_i^2\right) c_{2 \beta}
\left(-\frac{3}{2} e + \frac{2}{3} e^3 - \frac{37}{768} e^5 +
  \frac{11}{7680} e^7 \right),
\\
S_+^{(1)} &=& s_{2 \beta} \left(1 + c_i^2\right) \left(-\frac{3}{2} e
  + \frac{23}{24} e^3 + \frac{19}{256} e^5 + \frac{371}{5120} e^7
\right), \nonumber \\
\\
C_{\times}^{(1)} &=& s_{2 \beta} c_i \left(3 e - \frac{4}{3} e^3 +
  \frac{37}{384} e^5 - \frac{11}{3840} e^7\right),
\\
S_{\times}^{(1)} &=& c_{2 \beta} c_i \left(-3 e + \frac{23}{12} e^3 +
  \frac{19}{128} e^5 + \frac{371}{2560} e^7 \right),
\label{CS1end}
\eeq
and higher-order coefficients are listed in Appendix \ref{app:ccoeffs}.  This
expansion resembles that of~\cite{1995MNRAS.274..115M}, but it differs in that
we are here allowing for arbitrary binary inclinations via the angles
$(\iota,\beta)$ and we are consistently expanding to the same order in
eccentricity, without keeping prefactors of $(1 - e)^{-1}$. 
We have checked that our results in Eqs.~($3.7$)-($3.10$) and those in
Appendix \ref{app:ccoeffs} are consistent with Eqs.~($16$)-($18$) of
Ref.~\cite{1995MNRAS.274..115M}, after identifying their variable $\Theta$
with our $\iota$ and re-expanding their expressions in powers of $e$.

The maximum $k$ that one employs in the sums of the Bessel expansion,
$k_{\textrm{max}}$, is not generically equal to the maximum $\ell$ one uses in
the sums of the harmonic decomposition of the waveform,
$\ell_{\textrm{max}}$. This is because higher harmonics in the waveform, such
as $h_{+,\times} \sim \cos{3 \phi}$, can be re-expanded as powers of
$\cos{\phi}$ and $\sin{\phi}$ with standard trigonometric identities, which
then become powers of Bessel series via Eqs.~\eqref{moreno1}
and~\eqref{moreno2}. Cross-terms in the product of Bessel series combine to
produce harmonics of higher order. For example, if $k_{\textrm{max}}=4$, the
harmonic decomposition contains terms of the form $h_{+,\times} \sim h_{0}
e^{8} \cos{10 l}$. Increasing $k_{\textrm{max}}$ leads to terms that modify
$h_{0}$, but only up to $k_{\textrm{max}}=9$, beyond which $h_{0}$ is not
modified. Generically, this means that $\ell_{\textrm{max}} = k_{\textrm{max}}
+ 1 = N + 2$, so with our choices ($N = 8$, $k_{\textrm{max}} = 9$), we have
$\ell_{\textrm{max}} = 10$.

The expressions presented above depend on the eccentricity, which itself is a
function of time. Equation~\eqref{c0} can be solved for the orbital frequency
as a function of the eccentricity to obtain $F/F_0 =
[\sigma(e_0)/\sigma(e)]^{3/2}$, where $F_{0}$ is defined such that $F(e_{0}) =
F_{0}$.  This equation is not invertible for large eccentricities, but in the
limit $e \ll 1$ it yields:
\beq
\label{eccentr}
e &\sim& e_0 \chi^{-19/18} \left\{
1 
+  \frac{3323}{1824} e_0^2 \left[1 - \chi^{-19/9} \right]
+  \frac{15994231}{6653952} e_0^4 
\right. 
\nonumber \\
&& \left.
\left[1 - \frac{66253974}{15994231} 
\chi^{-19/9} + \frac{50259743}{15994231}
\chi^{-38/9} \right]  
\right.
\nonumber \\
&+& \left. \frac{105734339801}{36410425344} e_0^6 \left[
1 - \frac{1138825333323}{105734339801} \chi^{-19/9}
\right. \right. 
\nonumber \\
&+& \left. \left.
 \frac{2505196889835}{105734339801} \chi^{-38/9} 
- \frac{1472105896313}{105734339801} \chi^{-19/3}
\right] 
\right. 
\nonumber \\
&+& \left.
 {\cal{O}}(e_0)^8  \right\},
\eeq
where we have defined $\chi \equiv F/F_0$. Notice that Eq.~\eqref{eccentr} is a
series in odd powers of $e_{0}$, and as such, it possesses uncontrolled
remainders of ${\cal{O}}(e_{0}^{9})$.

\begin{figure}[htb]
  \includegraphics[width=8cm,clip=true]{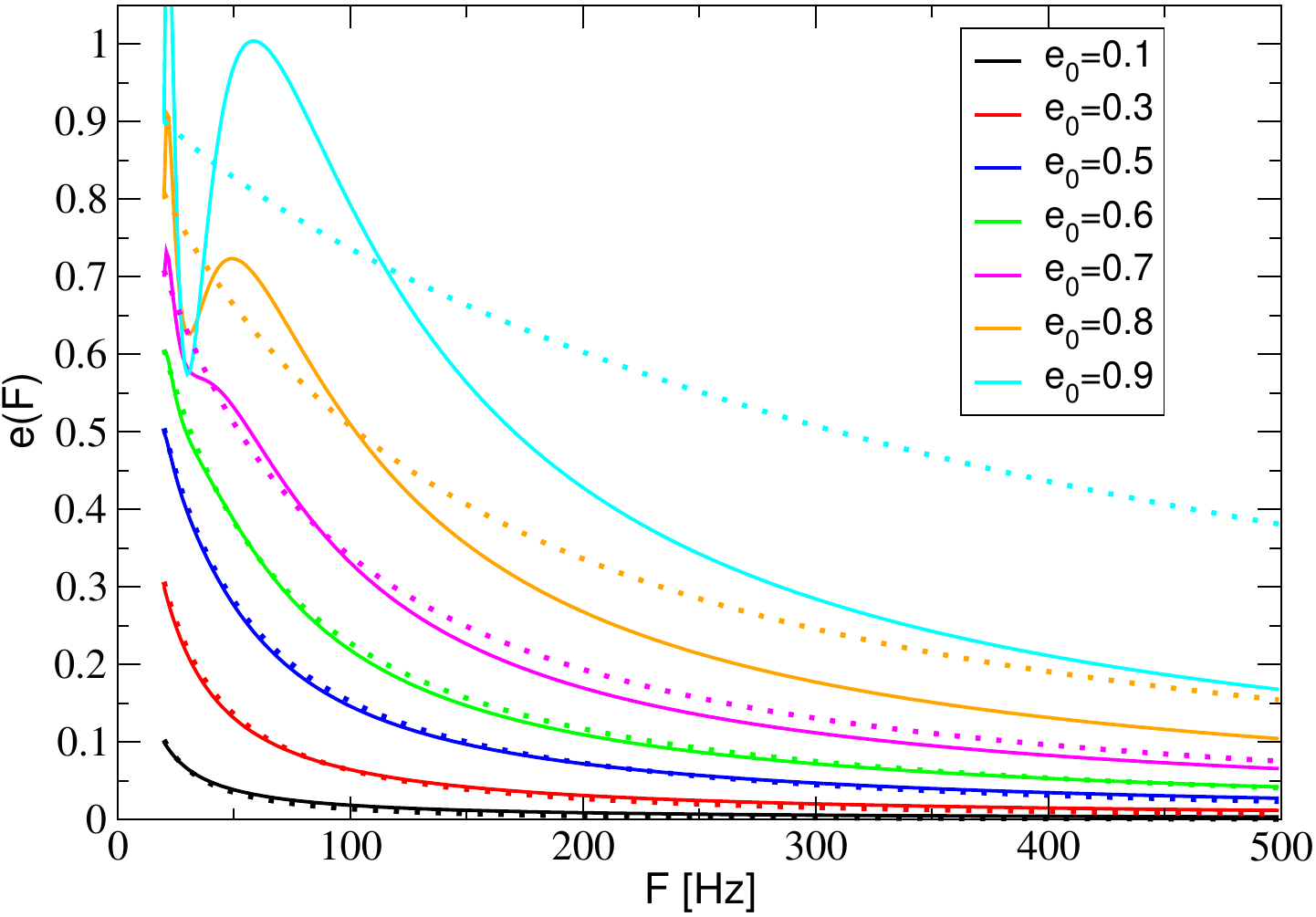}
\caption{\label{eofF} Eccentricity as a function of frequency for different
  value of the initial eccentricity $e_0$ evaluated at $F_{0} = 20$~Hz. Solid
  lines correspond to the eccentricity as given by Eq.~\eqref{eccentr} for
  $e_{0}=\{0.1,0.3,0.5,0.6,0.7,0.8,0.9\}$, in ascending order. Dotted lines
  correspond to the approximant of Eq.~\eqref{Pade} for the same initial
  eccentricities.}
\end{figure}
The inversion of the eccentricity as a function of frequency is not valid for
all frequencies and all initial eccentricities. Figure~\ref{eofF} plots
Eq.~\eqref{eccentr} as a function of frequencies for different initial
eccentricities, and initial frequency $F_{0}=20$~Hz.  Observe that for initial
eccentricities $e_{0} \leq 0.6$, $e(F)$ decays monotonically as a function of
frequency, as expected, but for $e_{0} > 0.6$ it ceases to be monotonic,
displaying two peaks.  This unphysical behavior is a signal that the expansion
in Eq.~\eqref{eccentr} breaks down. This occurs when the first correction in
the $e_{0}$ expansion ceases to be much less than unity.  This requirement
translates roughly to $F \gg 1.2 \; e_{0}^{18/19} F_{0}$, or simply $e_{0} \ll
0.8$.  We see then that truncating all expressions at $e < 0.4$ is consistent
with this requirement. More importantly, Fig.~\ref{eofF} shows that
Eq.~\eqref{eccentr} is well-behaved for this range of eccentricities.

Monotonicity, however, is not sufficient to guarantee that Eq.~\eqref{eccentr}
is valid up to $e = 0.4$. The ultimate test is to compare this power series
solution to the exact numerical inversion of Eq.~\eqref{c0}. We find that the
exact numerical solution can be fitted by the following phenomenological
fraction to better than $1 \%$ accuracy:
\be
e(F) = \frac{16.83 - 3.814 \; \beta^{0.3858}}{16.04 + 8.1 \; \beta^{1.637}},
\label{Pade}
\ee
where we have defined $\beta \equiv \chi^{2/3}/\sigma_{0}$. One can check that
in the limit $\chi \to 1$, Eq.~\eqref{Pade} equals $e_{0}$ with an accuracy of
roughly $1 \%$. Equation~\eqref{Pade} is plotted in Fig.~\ref{eofF} with
dotted lines.  Observe that the dotted lines are always close to the solid
lines for $e_{0} < 0.7$. The use of Eq.~(\ref{Pade}), however, would go
against the philosophy of this paper: to power-series expand all quantities in the
limit $e_{0} \ll 1$. We leave exploration of this type of resummation and
other for future work.

The waveform coefficients can now be written entirely as a generalized
power-series expansion in the frequency. Inserting Eq.~\eqref{eccentr} into
Eq.~\eqref{expansion} and re-expanding in the limit $e \ll 1$, we find, for
example
\begin{widetext}
\beq
C_+^{(2)} &=& \left(1 + c_i^2 \right) c_{{2\,\beta}} 
\left[ 
2 
- 5\,{{  e_0}}^{2}{{  \chi}}^{-{\frac {19}{9}}}
+ \left( -{\frac {16615}{912}} \,{{  \chi}}^{-{\frac {19}{9}}}+{\frac {19123}{912}}\,{{  \chi}}^{-{
\frac {38}{9}}} \right) {{  e_0}}^{4}
+ \left( -{\frac {8448925}{
207936}}\,{{  \chi}}^{-{\frac {19}{9}}}+{\frac {63545729}{415872}}\,{{
  \chi}}^{-{\frac {38}{9}}}
  \right. \right. 
  \nonumber \\
  &-& \left. \left.
  {\frac {234273299}{2079360}}\,{{  \chi}}^{
-{\frac {19}{3}}} \right) {{  e_0}}^{6} 
+
\left( -{\frac {41434504475}{568912896}}\,{{  \chi}}^{-{
\frac {19}{9}}}+{\frac {469672525907}{758550528}}\,{{  \chi}}^{-{
\frac {38}{9}}}-{\frac {778490172577}{632125440}}\,{{  \chi}}^{-{
\frac {19}{3}}}
\right. \right. 
\\ \nonumber 
&+& \left. \left.
{\frac {1559384621213}{2275651584}}\,{{  \chi}}^{-{
\frac {76}{9}}} \right) {{  e_0}}^{8} \right] +
{s_{{i}}}^{2} \left[  
{{  e_0}}^{2}{{  \chi}}^{-{\frac {19}{9}}} 
+
\left( {\frac {3323}{912}}\,{{  \chi}}^{-{\frac {19}{9}}}-{
\frac {1209}{304}}\,{{  \chi}}^{-{\frac {38}{9}}} \right) {{  e_0}}^{4}
+
\left( {\frac {1689785}{207936}}\,{{  \chi}}^{-{\frac {
19}{9}}}-{\frac {1339169}{46208}}\,{{  \chi}}^{-{\frac {38}{9}}}
\right. \right. 
\nonumber \\ \nonumber
&+& \left. \left.
{\frac {8690279}{415872}}\,{{  \chi}}^{-{\frac {19}{3}}} \right) {{  
e_0}}^{6}
+  \left( { 
\frac {8286900895}{568912896}}\,{{  \chi}}^{-{\frac {19}{9}}}-{\frac {
9897925427}{84283392}}\,{{  \chi}}^{-{\frac {38}{9}}}+{\frac {
28877797117}{126425088}}\,{{  \chi}}^{-{\frac {19}{3}}}
- {\frac {
1428551432057}{11378257920}}\,{{  \chi}}^{-{\frac {76}{9}}} \right) {{
  e_0}}^{8}  \right] .
\label{hardC+}
\eeq
\end{widetext}
Notice that in the limit $e_{0} \to 0$, Eq.~\eqref{hardC+} reduces to the 
appropriate circular limit: $C_{+} \to 2 (1 + c_{i}^{2}) c_{2 \beta}$.
As we can see, the modified coefficients are complicated and
unilluminating, which is why we do not present the remaining ones
here. Nonetheless, it is straightforward to insert Eq.~\eqref{eccentr}
into the waveforms of Eq.~\eqref{expansion} to obtain amplitude
corrections as a function of the orbital frequency. 

\section{Fourier Transform of the Waveform in the SPA} 
\label{Sec:SPA}

In this section we calculate the Fourier transform of the waveform computed in
the previous section. To do so, we shall employ the SPA~(see \cite{bender},
and \cite{Cutler:1994ys} for a discussion in the context of GW data analysis),
which is an expansion in the ratio of the radiation-reaction time scale to the
orbital period.  In this asymptotic expansion we need only keep the
controlling factor, since subdominant terms can in general be neglected for
matched-filtering purposes~\cite{Droz:1999qx}.  Recently, it has been proposed
that amplitude corrections in the waveforms might play a critical role in the
data analysis problem~\cite{SinVecc00a,SinVecc00b,MH02,HM03,
  ChrisAnand06,ChrisAnand06b, Arun:2007qv,Arun:2007hu,TriasSintes07,
  PorterCornish08,ABFO08}, but we defer a discussion of those corrections to
future work. We shall here primarily follow the prescription
of~\cite{Krolak:1995md}.

Let us begin by reviewing the SPA, following Ref.~\cite{bender}. Consider the
generalized Fourier integral
\be
\label{gen-SPA}
I(y) = \int_a^b g(t) e^{i y \psi(t)} dt, 
\ee
where $g(t)$, $\psi(t)$, $a$, $b$ and $y$ are all real. In order to find the
asymptotic behavior of such an integral as $y \to + \infty$, one searches for
{\emph{stationary}} points, namely those where $\dot{\psi} = 0$. This is
because in the neighborhood of stationary points the integrand oscillates less
rapidly, and there is less cancellation between adjacent
subintervals~\cite{bender}. Thus, the asymptotic behavior of
Eq.~(\ref{gen-SPA}) is given by
\be
\label{gen-SPA2}
I(y) \sim g(a) \; e^{i y \psi(a) \pm i \pi/(2 p)} \left[ \frac{p!}{y
    |\psi^{(p)}(a)|}\right]^{1/p} \frac{\Gamma(1/p)}{p}
\ee
as $y \to + \infty$, where $\psi^{(p)}(a)$ stands for the $p$-th derivative of
$\psi$ evaluated at $t = a$. In Eq.~\eqref{gen-SPA2} the factor of $e^{\pm i
  \pi/(2 p)}$ has a positive sign if $\psi^{(p)}(a)>0$, and a negative sign if
$\psi^{(p)}(a)<0$.  Here we have chosen the stationary point to be located at
$t=a$, such that $\psi^{(1)}(a) = \ldots = \psi^{(p-1)}(a) = 0$.
  
Let us then define the Fourier transform of some time-series $B(t)$ as
\be
\tilde{B}(f) \equiv \int_{-\infty}^{\infty} B(t) e^{2 \pi i f t} dt,
\ee
and let us write the time-domain waveform as the product of a slowly-varying
amplitude ${\cal{A}}(t)$ and a rapidly-varying cosine with phase $\ell
\phi(t)$ and $\ell >0$. Then, the Fourier transform of the cosine (denoted by
a subscript $C$) becomes
\be
\label{fft-SPA1}
\tilde{B_C}(f) = \frac{1}{2} \int_{-\infty}^{\infty} {\cal{A}}(t)
\left(e^{2 \pi i f t + i \ell \phi(t)}  + e^{2 \pi i f t - i \ell \phi(t)}
\right) dt. 
\ee
The first term in Eq.~(\ref{fft-SPA1}) does not contain any stationary points
and, thus, it vanishes via the Riemann-Lebesgue lemma~\cite{bender}. The
second term, however, does have a stationary point at the value $t_0$ where
$\ell \dot{\phi} (t_0) = 2 \pi f$, which defines the {\emph{stationary phase
    condition}}: $F(t_0) = f/\ell$, where we have defined $F(t) \equiv
\dot{\phi}(t)/(2\pi)$. Thus, the asymptotic behavior of the Fourier transform
of a cosine time-series is
\be
\label{fft-SPA2}
\tilde{B}_C(f) = \frac{{\cal A}(t_0)}{2\sqrt{\ell \dot F(t_0)}} 
e^{-i \left(\Psi + \pi/4 \right)},
\ee
where we have defined the phase
\be
\Psi : = - 2 \pi f t_0 + \ell \phi(t_0).
\ee
The Fourier transform of a sine times-series is then simply $\tilde{B}_S(f) = i
\tilde{B}_C(f)$.  Equation~\eqref{fft-SPA2} is identical to
Eq.~(\ref{gen-SPA2}) with $p=2$, $g(t) = {\cal{A}}(t)/2$, $y = f$ and
$\psi(t)= 2 \pi t - \ell \phi(t)/f$.  In obtaining this solution we have
implicitly assumed that $d( \ln {\cal{A}})/dt \ll d\phi/dt$ and $d^2\phi/dt^2
\ll \left(d\phi/dt\right)^2$, which mathematically enforces the physical
condition that the amplitude varies much more slowly than the phase.  In
Eq.~(\ref{fft-SPA2}) there is an extra factor of two relative to
Eq.~(\ref{gen-SPA2}), because the phase $\Psi$ is not monotonic in its
range\footnote{The stationary phase integral can be broken down into two
  parts, inside each of which $\psi$ is monotonic. Each of these integrals has
  a stationary phase contribution that leads to the factor of two in
  Eq.~\eqref{gen-SPA2}.  See~\cite{Droz:1999qx} for more details.}.

In order to find the full solution, we must solve for the phase $\Psi$.
Defining the quantity $\tau \equiv F/\dot{F}$, we can rewrite $\phi(F)$ and
$t(F)$ as
\be
\label{PhiF}
\phi(F) = 2 \pi \int^F \tau' \; dF',
\qquad
t(F) = \int^F \frac{\tau'}{F'} dF',
\ee
which then leads to 
\be
\Psi[F(t_{0})] = 2 \pi \int^{F(t_{0})} \tau' \left(\ell - \frac{f}{F'}\right) dF'. 
\label{new-kostas}
\ee
Of course, these expressions must be evaluated at the stationary point,
given above by $F(t_{0}) = f/\ell$. 

\subsubsection{Circular Case}
The above formalism can be understood better by studying the well-known
circular case. Let us then solve explicitly for the Fourier transform of the
response function $h(t)$ in the SPA for a binary in circular orbit
($e=0$). The response function is defined via the linear combination
\be
\label{circ-response}
h(t) = F_+(\theta_S\,,\phi_S\,,\psi_S) h_+ + 
F_{\times}(\theta_S\,,\phi_S\,,\psi_S) h_{\times},
\ee
where $F_+$ and $F_{\times}$ are the so-called beam-pattern functions that
characterize the response of the detector to an impinging GW and are also
slowly-varying (see e.g.~\cite{BBW05a}). The orbital frequency evolution is
given by
\be
\frac{dF}{dt} = \frac{48}{5\pi {\cal M}^2} 
\left(2\pi {\cal M} F\right)^{11/3}
\ee
(see e.g.~the leading-order contribution to Eq.~(A.2) in
\cite{ChrisAnand06}). We can rewrite the response function of
Eq.~\eqref{circ-response} as $h(t) = h_C(t) + h_S(t)$, where
\beq
h_C(t) &=& {\cal{A}} \; Q_C(\iota,\beta) \; \cos{2 \phi},
\\
h_S(t) &=& {\cal{A}} \; Q_S(\iota,\beta) \; \sin{2 \phi},
\eeq
and where the amplitude ${\cal A}$ is a function of frequency, defined in
Eq.~\eqref{amplitude}. We have here introduced the following functions of the
polarization and inclination angles:
\begin{subequations}
\begin{eqnarray}
Q_C(\iota,\beta) 
&\equiv& 
2 \left(1 + c_i^2\right) c_{2 \beta} F_+ - 4 c_i s_{2 \beta} F_{\times},\\
Q_S(\iota,\beta) 
&\equiv& 
2 \left(1 + c_i^2\right) s_{2 \beta} F_+ + 4 c_i c_{2 \beta} F_{\times}.
\end{eqnarray}
\end{subequations}

The Fourier transform can then be computed in the SPA via
Eq.~(\ref{fft-SPA2}) with $\ell = 2$. We thus obtain
\beq
\tilde{h}_{C}(f) &=& - \left(\frac{5}{384}\right)^{1/2} \pi^{-2/3} 
\frac{{\cal{M}}^{5/6}}{D_L} \; Q_C(\iota,\beta) \; 
\left[2F(t_0)\right]^{-7/6} 
\nonumber \\
&\times&
e^{-i \left(\Psi + \pi/4 \right)}. 
\eeq
We can solve for the time and phase functions to obtain
\begin{eqnarray}
\phi(F) &=& \phi_c + 2 \pi \int^F \tau' dF' 
= \phi_c - \frac{1}{32} \left(2 \pi {\cal{M}} F\right)^{-5/3},
\nonumber \\
t(F) &=& t_c + \int^F \f{\tau'}{F'} dF' 
= t_c - \frac{5 {\cal M}}{256} \left(2 \pi {\cal M}
  F\right)^{-8/3},  
\nonumber \\
\end{eqnarray}
where $\phi_c$ and $t_c$ are the orbital phase and time of coalescence.
Substituting the stationary phase condition $F(t_0) = f/2$ into these
expressions, the phase $\Psi$ becomes
\be
\Psi = -2 \pi f t_{c} + \bar\phi_{c} - \frac{3}{128 x},
\ee
where we have defined $x \equiv (\pi {\cal{M}} f)^{5/3}$, and $\bar{\phi}_{c}$
is the GW phase at coalescence.  The argument of the exponential of the
Fourier transform is then
\be
-i \ln \left[\frac{\tilde{h}_{\rm (circ)}}{|\tilde{h}_{\rm (circ)}|}
\right] = 2 \pi f t_c  -  \bar\phi_c - \frac{\pi}{4}  
+ \frac{3}{128} \left( \pi {\cal{M}} f\right)^{-5/3},  
  \label{circphase}
\ee
and the full Fourier transform becomes
\beq
\tilde{h}_{\rm (circ)} &=& - \left(\frac{5}{384}\right)^{1/2} \pi^{-2/3} 
\frac{{\cal{M}}^{5/6}}{D_L} \; Q(i,\beta) \; f^{-7/6} 
\\ \nonumber 
&\times&
\exp\left[
  {i \left(2 \pi f t_c  - \bar\phi_c - \frac{\pi}{4}  + \frac{3}{128} 
      \left( \pi {\cal{M}} f\right)^{-5/3} \right)} \right],
\eeq
where $Q = Q_{C} + i Q_{S}$.  Equation~\eqref{circphase} is in agreement with
well-known results in the literature \cite{Cutler:1994ys}, when we keep in
mind that the GW frequency $f = F_{\GW} = 2 F$ is usually adopted to write
down all results in calculations involving circular binaries.

\subsubsection{Eccentric Case}
\label{ecc-case-FD}

Let us now focus on eccentric inspirals. Once more, we must consider the
response function of the detector, defined by Eq.~\eqref{circ-response},
except that now $h_{+,\times}$ correspond to the eccentric waveforms discussed
in Sec.~\ref{Sec:model}.  For eccentric waveforms, the response function
becomes
\be
\label{response}
h(t) = {\cal{A}} \sum_{\ell=1}^{10} 
\left[ \Gamma_\ell \cos{\left(\ell \; l \right)} + 
\Sigma_\ell \sin{\left(\ell \; l \right)} \right],
\ee
where $\cal{A}$ is a function of frequency defined in Eq.~(\ref{amplitude}),
and where we have defined
\be
\label{Gamma-Sigma}
\Gamma_{\ell} \equiv F_+ C_+^{({\ell})} + F_{\times}  C_{\times}^{({\ell})},
\qquad
\Sigma_{\ell} \equiv F_+ S_+^{({\ell})} + F_{\times} S_{\times}^{({\ell})}.
\ee
These coefficients are slowly-varying functions of time, which can be written
as functions of the orbital frequency for some given initial eccentricity
$e_0$.  By using the trivial trigonometric identity $\cos(\ell \;
l+\phi)=\cos(\ell \; l)\cos(\phi)-\sin(\ell \; l)\sin(\phi)$ we can combine
terms into a single sum of the form:
\be
h(t) = {\cal{A}} \sum_{\ell=1}^{10} 
\alpha_\ell \cos{\left(\ell \; l +\phi_\ell\right)},
\ee
where
\begin{subequations}
\begin{eqnarray}
\label{alphak}
\alpha_\ell&=&{\rm sign}(\Gamma_\ell)\sqrt{\Gamma_\ell^2+\Sigma_\ell^2},\\
\phi_\ell&=&\tan^{-1}\left(-\f{\Sigma_\ell}{\Gamma_\ell}\right).
\label{phik}
\end{eqnarray}
\end{subequations}
We shall not present the coefficients $\alpha_\ell$ and $\phi_\ell$ here, but
they can be straightforwardly calculated using results from the previous
section.

The Fourier transform in the SPA then becomes 
\beq
\label{SPA-FFT-eccen}
\tilde{h} &=& - \left(\frac{5}{384}\right)^{1/2} \pi^{-2/3}
\frac{{\cal{M}}^{5/6}}{D_L} \left[2 F(t_0)\right]^{-7/6} 
\nonumber \\
&\times&
\frac{\left(1 -
    e^2\right)^{7/4}}{\left(1 + \frac{73}{24} e^2 + 
    \frac{37}{96} e^4 \right)^{1/2}} 
\\ \nonumber 
&\times&
\sum_{\ell=1}^{10} \alpha_\ell \sqrt{\frac{2}{\ell}}  e^{-i\phi_\ell[F(t_0)]}
e^{-i \left(\Psi + \pi/4\right)},   
\eeq
where we have used the fact that for eccentric orbits the 
orbital phase evolution is given by~\cite{Peters:1964zz}
\be
\label{fdotecc}
\frac{dF}{dt} = \frac{da}{dt} \frac{dF}{da} = 
\frac{48}{5 \pi {\cal{M}}^{2}} \left(2 \pi {\cal{M}} F\right)^{11/3}
\frac{\left(1 + \frac{73}{24} e^2 + \frac{37}{96} e^4 \right)}{\left(1 - e^2\right)^{7/2}}.
\ee
The frequency that appears in Eq.~\eqref{SPA-FFT-eccen} can be thought of as 
an {\emph{instantaneous mean orbital frequency}}. Such a quantity is ``instantaneous'' 
in the sense that it evolves on a radiation-reaction timescale. 
The factor of $1/\sqrt{\ell}$ comes about due to the factor of
$(d^2\phi/dt^2)^{-1/2}$ in Eq.~(\ref{fft-SPA2}). Also note that now $\phi(t_0)
= \ell l(t_0)$ and the factor of $\phi_\ell$ cannot be pulled out of the sum
because it depends on $\ell$. One can check that in the limit $e_{0} \to 0$,
$\alpha_{2} e^{-i \phi_{2}} \to Q$ and we recover the circular limit.

The phase $\Psi$ must be evaluated at the stationary point $t_{0}$, which
is here defined implicitly via $\ell \dot{l}(t_0) = 2 \pi f$ or simply $F(t_0)
= f/\ell$, as already discussed. The phase $\Psi$ is then essentially
Eq.~\eqref{new-kostas}, which requires knowledge of the characteristic time
scale $\tau$. Unlike the circular case, for eccentric inspirals the integral
over $\tau$ can only be done approximately, since
\be
\tau = \frac{5 {\cal{M}}}{96} \left(2\,\pi {\cal{M}} F\right)^{-8/3}
\frac{\left(1 - e^2\right)^{7/2}}{1 + \frac{73}{24} e^2 + \frac{37}{96}
e^4},
\label{tau1}
\ee
and $e$ is a slowly varying function of $F$ that cannot be inverted in closed
form.  We can achieve this inversion asymptotically for small eccentricities,
by first expanding Eq.~\eqref{tau1} in $e \ll 1$
\beq
\tau &\sim& \frac{5 {\cal{M}} }{96} \left(2\,\pi {\cal{M}}
  F\right)^{-8/3} \left[ 1 - \frac{157}{24} e^2 + \frac{13759}{576}
  e^4 
\right. 
\nonumber \\
&-& \left.
 \frac{999793}{13824} e^6 +  \frac{70021111}{331776} e^{8}
 + {\cal{O}}(e^{10}) \right].
\eeq
Since the eccentricity as a function of frequency is given by
Eq.~\eqref{eccentr}, the characteristic time becomes
\beq
\tau &\sim& \frac{5 {\cal{M}}}{96} \left(2\,\pi {\cal{M}} F\right)^{-8/3}
\left\{ 
1 
- \frac{157}{24} e_0^2 \chi^{-19/9} 
\right.
\nonumber \\
&+& \left. 
 e_0^4 \left[ \frac{1044553}{21888} \chi^{-38/9}
  - \frac{521711}{21888} \chi^{-19/9} \right]
\right. 
\nonumber \\
&+& \left. 
 e_0^6 \left[ \frac{3471049619}{9980928}
\chi^{-38/9} - \frac{265296245}{4990464}
\chi^{-19/9} 
\right. \right. 
\nonumber \\
&-& \left. \left. 
 \frac{135641025}{369664}
\chi^{-19/3} \right] 
+ e_{0}^{8} \left[
-\frac{450735126075}{112377856} \chi^{-19/3} 
\right. \right. 
\nonumber \\
&+& \left. \left.
 \frac{25654857812777}{18205212672} \chi^{-38/9}
+ \frac{158823466804555}{54615638016} \chi^{-76/9}
\right. \right. 
\nonumber \\
&-& \left. \left.
  \frac{1301043440515}{13653909504} \chi^{-19/9} 
\right]
+ {\cal{O}}(e_0^{10})
\right\},
\eeq
where as usual $\chi \equiv F/F_{0}$. This is a generalization
of Eq.~(A8) of Ref.~\cite{Krolak:1995md} to higher powers of eccentricity. 

We can now compute the new phase [Eq.~(\ref{new-kostas})] by integrating the
characteristic time. Using the stationary phase condition, $F(t_0) = f/\ell$,
we obtain
\beq
\label{final-new-Kostas}
\Psi_{\ell} &=& \ell \phi_c - 2 \pi f t_c - \frac{3}{128 \; x} \left({\ell\over2}\right)^{8/3} \left[1 -
  \frac{2355}{1462} e_0^2 \chi^{-19/9} 
\right.
\nonumber \\
&+& \left. 
 e_0^4 \left(\frac{5222765}{998944}
    \chi^{-38/9} - \frac{2608555}{444448} \chi^{-19/9} \right) 
\right.
\nonumber \\
&+& \left. 
e_0^6 \left( 
    - \frac{75356125}{3326976} \chi^{-19/3} - \frac{1326481225}{101334144}
    \chi^{-19/9}
\right. \right.
\nonumber \\
&+& \left. \left.
 \frac{17355248095}{455518464} \chi^{-38/9} \right) 
 + e_{0}^{8} \left(
 - \frac{250408403375}{1011400704} \chi^{-19/3}
 \right. \right. 
 \nonumber \\
 &+& \left. \left.
 \frac{4537813337273}{39444627456} \chi^{-76/9}
 - \frac{6505217202575}{277250217984} \chi^{-19/9}
 \right. \right. 
 \nonumber \\
 &+& \left. \left.
 \frac{128274289063885}{830865678336} \chi^{-38/9} \right)
 + {\cal{O}}(e_{0}^{10}) \right],
\eeq
where we recall that $x \equiv (\pi {\cal{M}} f)^{5/3}$, and $\Psi$ has now
become a function of $\ell$.  
We have checked that the first few terms in the phase 
of Eq.~\eqref{final-new-Kostas} agree with the phase computed in Eq.~$({\textrm{A}}10)$ 
of~\cite{Krolak:1995md}.
Notice that when we apply the stationary phase
condition to $e(F)$ we must also rescale $F_{0} \to f_{0}/\ell$, so that
$e(f_{0}) = e_{0}$.  Otherwise, the eccentricity function would not be
properly normalized.

Combining all pieces together we obtain the Fourier transform in the SPA,
namely
\be
\label{fft}
\tilde{h} = \tilde{{\cal{A}}} f^{-7/6} \sum_{\ell=1}^{10}
\xi_\ell \; \left(\frac{\ell}{2}\right)^{2/3}  
\; e^{-i \left(\pi/4 + \Psi_\ell \right)},
\ee
where we have defined 
\beq
\tilde{{\cal{A}}} &=& - \left(\frac{5}{384}\right)^{1/2} \pi^{-2/3}
\frac{{\cal{M}}^{5/6}}{D_L},
\\
\xi_\ell &=& \frac{\left(1 -
    e^2\right)^{7/4}}{\left(1 + \frac{73}{24} e^2 + 
  \frac{37}{96} e^4\right)^{1/2}} \; \alpha_\ell \; e^{-i\phi_\ell(f/k)}.
\label{def:xik}
\eeq
This is the Fourier transform of the waveform for eccentric inspirals in the
SPA. Note that we have kept up to ten harmonics, which corresponds to a
consistent expansion in the eccentricity to ${\cal{O}}(e^8)$ both in the
amplitude and in the phase. We already saw in Sec.~\ref{Sec:kepler} that this
is enough to model the Bessel function to high accuracy even for relatively
high eccentricities.

The Fourier transform presented here depends on the coefficients $\xi_\ell$
that need to be {\emph{re-expanded}} in the limit $e_{0} \ll 1$. These
coefficients can be obtained from Eq.~\eqref{def:xik}, using the definition of
$e(F)$ in Eq.~\eqref{eccentr}, $\alpha_{\ell}$ in Eq.~\eqref{alphak},
$\phi_{\ell}$ in Eq.~\eqref{phik} and $\Gamma_{\ell}$ and $\Sigma_{\ell}$ in
Eq.~\eqref{Gamma-Sigma}, where $C_{+,\times}$ and $S_{+,\times}$ are given in
Appendix~\ref{app:ccoeffs}.  The resulting expression must then be re-expanded
in the limit $e_{0} \ll 1$ to ${\cal{O}}(e_{0}^{8})$.  We shall not present
these expressions here in full generality, since they are lengthy and
complicated.  Instead we present partial results for $\xi_{k}$ as a function
of $e(F)$ in Appendix \ref{app:xik} for an optimally oriented binary ($\iota =
\beta = 0$). In the next section, we shall employ these expressions in
combination with Eq.~\eqref{eccentr}, and re-expand them in $e_{0} \ll 1$ to
eighth order to compute the SNR.

\section{SNR calculation} 
\label{Sec:SNR}

In this section we compute the SNR using the Fourier transform of the waveform in the SPA
[Eq.~(\ref{fft})]. The SNR is defined via 
\be
\rho^2 \equiv 4 \Re \int_{f_{\rm low}}^{f_{\rm high}} \frac{ \tilde{h} \;
  \tilde{h}^{\star} }{S_n(f)} df,
  \label{SNR-form}
\ee
where $S_n(f)$ is the one-sided noise power spectral density and the star
superscript stands for complex conjugation. The noise curves of the AdvLIGO, ET
and LISA detectors are taken from Refs.~\cite{AISS05}, \cite{ETPSD} and
\cite{BBW05a} respectively; for LISA, in particular, we adopt the simple
``angle-averaged'' model discussed in~\cite{BBW05a}.
\subsection{Limits of Integration}

The upper frequency of integration, $f_{\rm high}$, is either the frequency at
which the motion transitions from inspiral to plunge or the maximum frequency
at which the detector noise is under control. Since the noise power spectral
densities for LIGO and ET increase steeply at high frequency, we will choose 
\begin{eqnarray}
f_{\rm high}^{\LISA} &=& \min\left[2 F_{\ISCO},1 \; {\rm{Hz}}\right],
\\
f_{\rm high}^{\LIGO} &=& f_{\rm high}^{\ET} = 2 F_{\ISCO}.
\end{eqnarray}
In the previous equations, as customary in the GW literature, we (somewhat
arbitrarily) pick the ISCO frequency to be $F_{\ISCO} \equiv 6^{-3/2} (2 \pi
M)^{-1}$, in analogy with the orbital frequency of a test particle at the ISCO
of the Schwarzschild spacetime. In Appendix \ref{app:ISCO} we discuss possible
eccentricity-induced modifications to this conventional ISCO frequency,
concluding that such modifications should not introduce significant
corrections to our SNR calculations.

The lower limit of integration $f_{\rm low}$ is determined by a seismic (or
acceleration) noise cut-off:
\begin{eqnarray}
f_{\rm low}^{\LIGO} &=& f_{s}^{\LIGO} = 20 \; {\textrm{Hz}},
\nonumber \\
f_{\rm low}^{\ET} &=& f_{s}^{\ET},
\nonumber \\
f_{\rm low}^{\LISA} &=& f_{\rm acc} = 10^{-4} \; {\textrm{Hz}},
\end{eqnarray}
where we shall investigate the ET SNR with $f_{s}^{ET} = 1$~Hz or $f_{s}^{ET}
= 10$~Hz. The quantity $f_{\rm acc}$ corresponds to the minimum frequency at
which acceleration noise is under control in LISA.

Different harmonic components will generically sample different frequency
ranges if we terminate all integrations when the dominant quadrupole GW
frequency equals the ISCO frequency. To ensure that higher harmonics do not
exceed the region of validity, following~\cite{Arun:2007qv}, we shall truncate
the waveforms with unit step functions $\Theta(x)$ ($\Theta(x)=1$ if $x\geq 0$
and zero otherwise):
\begin{eqnarray}
\tilde{h}_{\LIGO/\ET} &=& \tilde{{\cal{A}}} f^{-7/6} \sum_{\ell=1}^{10}
\left(\frac{\ell}{2}\right)^{2/3}  \xi_\ell \; e^{-i \Psi_\ell }
\nonumber \\
&\times&  \Theta\left(\ell  f_{\rm high}^{\LIGO/\ET} - 2 f\right) ,\label{eq:AL-ET}
\end{eqnarray}
where we have removed the factor of $\pi/4$ in the phase, since it cancels out
in SNR calculations. The step function guarantees that higher harmonics are
truncated at the correct upper frequency cut-off. 

LISA sources can spend several years in the LISA band, an issue that must be
accounted for, since the detector will not take data for more than a few
years.  Following~\cite{BBW05a,Arun:2007qv}, we shall multiply the waveform by
an additional step function:
\begin{eqnarray}
\tilde{h}_{\LISA} &=& \frac{\sqrt{3}}{2} \tilde{{\cal{A}}} f^{-7/6} \sum_{\ell=1}^{10}
\left(\frac{\ell}{2}\right)^{2/3}  \xi_\ell \; e^{-i \Psi_\ell } 
\nonumber \\
&\times& \Theta\left(\ell  f_{\rm high}^{\LISA} - 2 f\right) \Theta(2 f - \ell f_{\rm yr}^{\LISA}),
\end{eqnarray}
where $f_{\rm yr}$ is the GW frequency of the fundamental harmonic at a time
$T$ before the system reaches the ISCO (see Appendix~\ref{Fyr-app} for a
discussion of how to calculate this quantity for eccentric inspirals).  We
shall here choose $T$ to be equal to one year (hence assuming, somewhat
optimistically, that we can observe the whole last year of inspiral). This
step-function cut-off guarantees that all harmonics are integrated for no more
than one year, which is the higher-harmonic generalization of the criterion
used in~\cite{BBW05a}. Note also that we have multiplied the LISA waveform
amplitude by a geometrical correction factor of $\sqrt{3}/2$ (see
\cite{Cutler98,BBW05a} for details).

With these considerations in mind, the SNR is given by
\be
\rho^2_{A} = 4 \Re 
\int_{f_{\rm low}^{A}}^{\ell_{\rm max} f_{\rm high}^{A}} 
\frac{ \tilde{h}_{A} \; \tilde{h}^{\star}_{A} }{S_n^{A}}  df,
\ee
where $A$ stands for any of LIGO, ET or LISA.  Caution should be exercised in
comparing results between different detectors. Even for astrophysical systems
with the same masses, different detectors have different low-frequency
cut-offs, and the initial eccentricity $e_0$ is defined as the value of $e$ at
that frequency.  For example, a 100 $M_\odot$ system with $e_0=0.3$ does not
correspond to the same astrophysical system when we discuss AdvLIGO, whose
seismic cut-off is $20$~Hz, and when we discuss ET, whose seismic cut-off is
$10$~Hz or $1$~Hz.

\subsection{Results}

\begin{figure*}[htb]
\begin{center}
\begin{tabular}{cc}
  \includegraphics[width=8cm,clip=true]{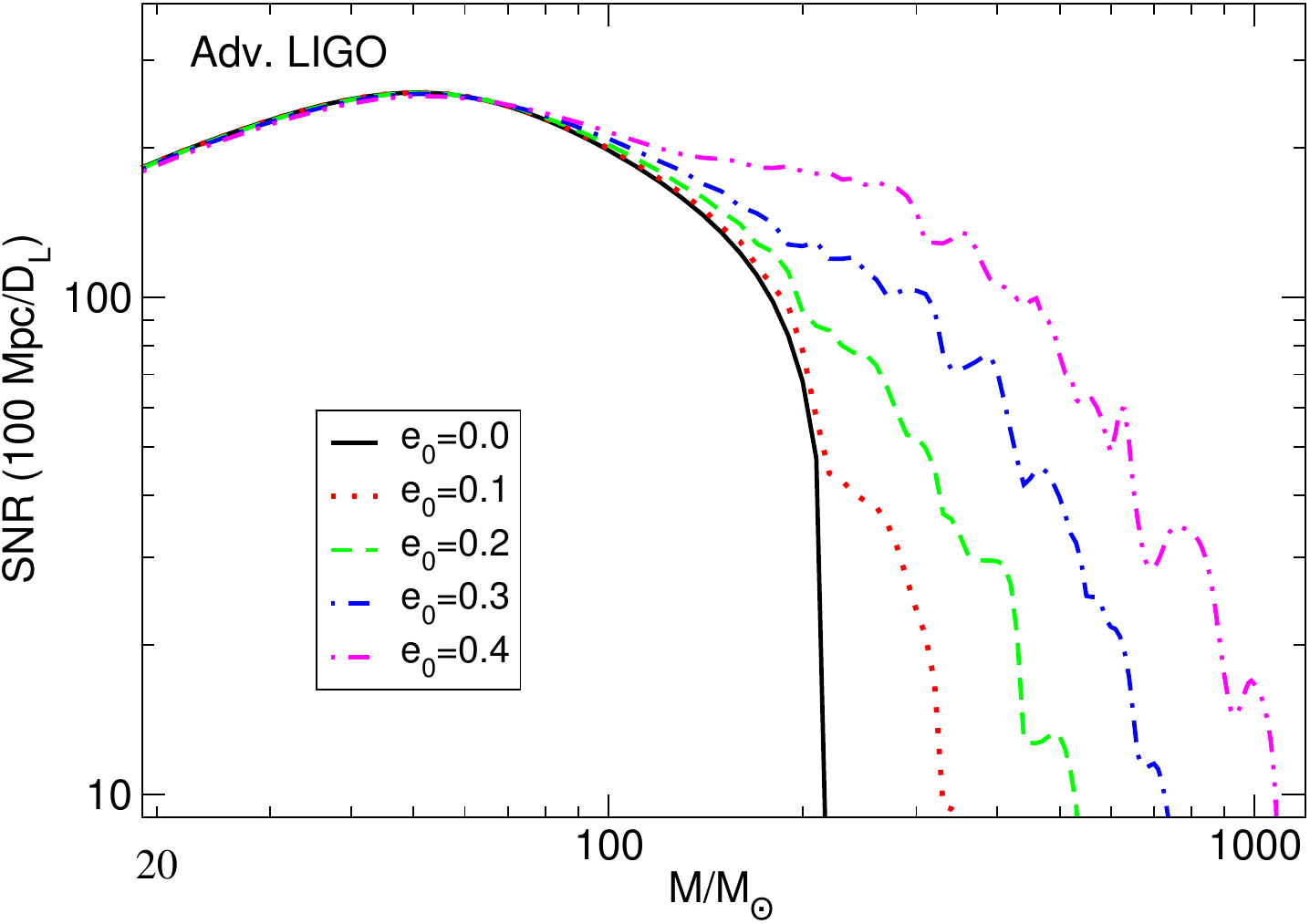}
   \includegraphics[width=8cm,clip=true]{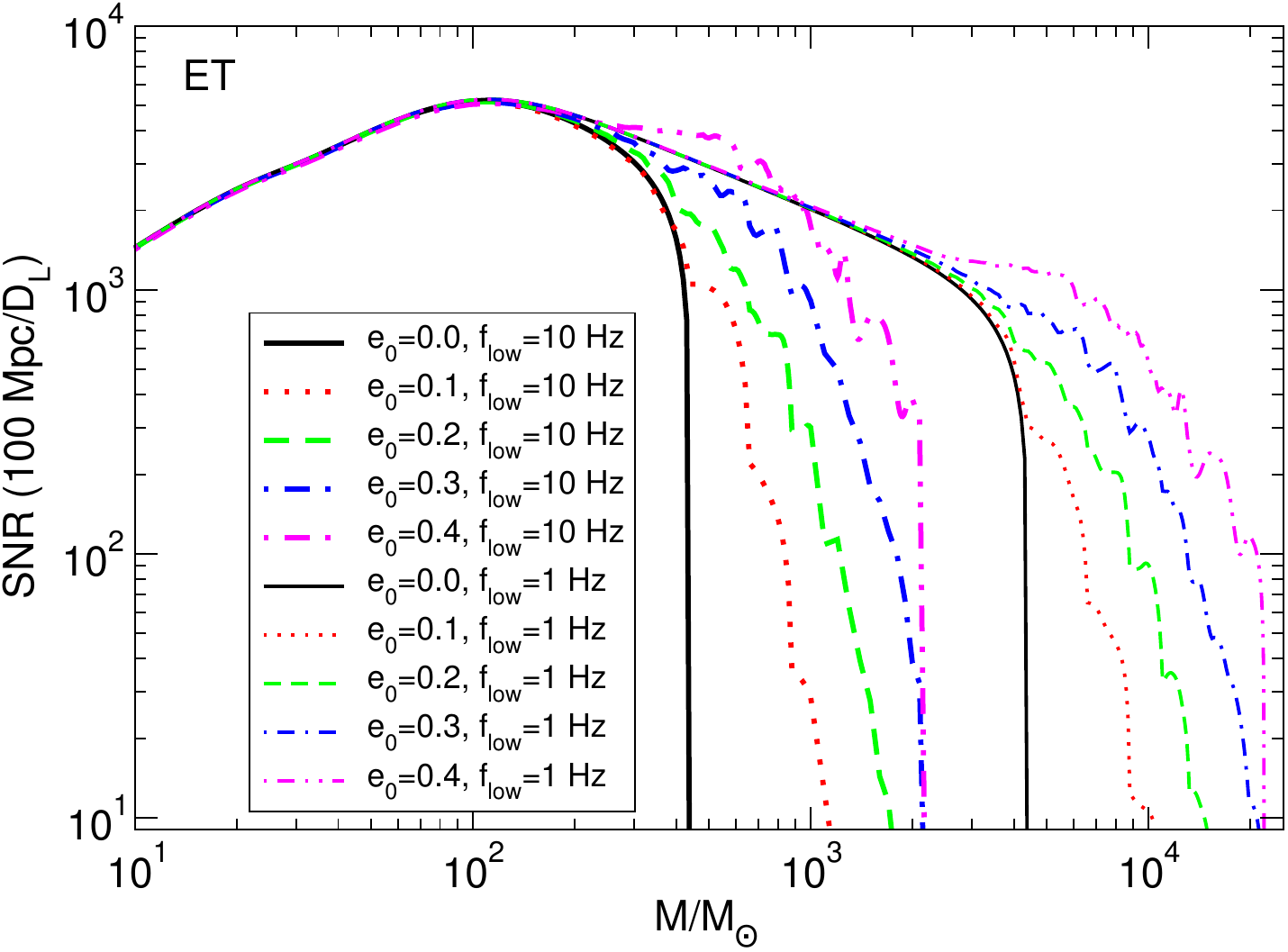} \\ 
    \includegraphics[width=8cm,clip=true]{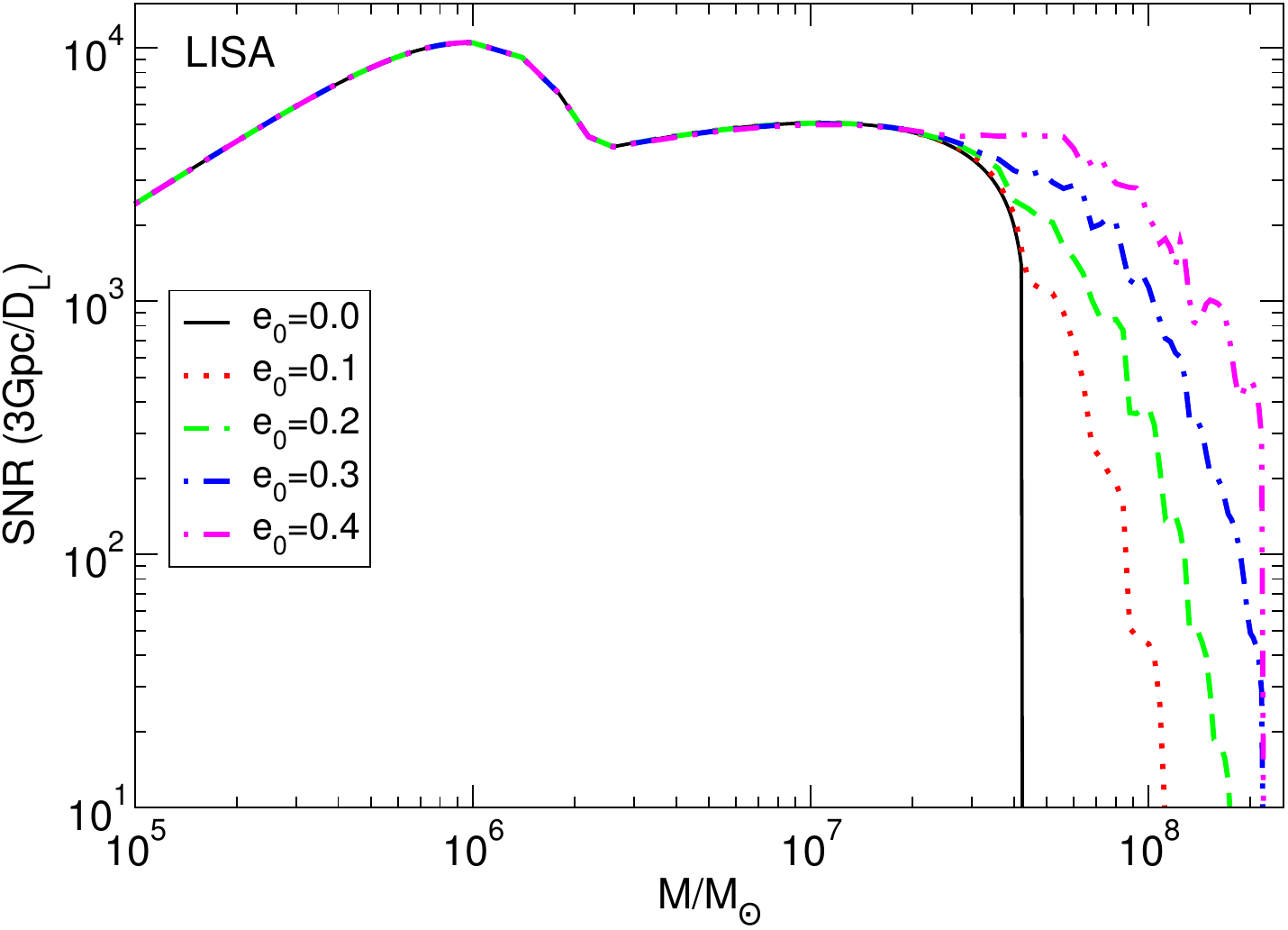}
\end{tabular}
\end{center}
\caption{\label{SNR} SNR for an equal-mass binary at optimal orientation as a
  function of total mass in solar mass units for different initial
  eccentricities.  The top figures correspond to the
  AdvLIGO (left) and ET (right) detectors, while results for the LISA detector 
  are shown in the bottom panel.}
\end{figure*}

Figure~\ref{SNR} plots the SNR for an equal-mass system as a function of the
total binary mass expressed in solar mass units. This SNR is computed at the
optimal binary orientation $(\iota=\beta=\theta_S=\phi_s=\psi_S=0)$.  As
discussed earlier, for each detector the initial eccentricity is computed at
some (somewhat conventional) lowest cut-off frequency, which is different for
each detector. We recall that for AdvLIGO and ET this lower cut-off
corresponds to the seismic noise ``wall'' ($20$~Hz and either $1$ or $10$~Hz),
while for LISA we (conservatively) adopt a lower cut-off at $10^{-4}$~Hz.

As a generic trend, the mass reach increases by as much as a factor of five
for the largest initial eccentricities explored here. For systems with $e_{0}
= 0.4$ AdvLIGO could observe binaries with total mass up to $10^{3} M_{\odot}$,
ET could see systems up to $2 \times 10^{3} M_{\odot}$ or $2 \times 10^{4}
M_{\odot}$ (for a $10$ or a $1$~Hz low-frequency cut-off, respectively), and
LISA could see binaries up to $2 \times 10^{8} M_{\odot}$. This is to be
compared to circular inspiral mass reaches of approximately $200 M_{\odot}$
for AdvLIGO, $400 M_{\odot}$ or $4 \times 10^{3} M_{\odot}$ (for a $10$ or a
$1$~Hz low-frequency cut-off) and $4 \times 10^{7} M_{\odot}$ for LISA.

Another feature of this figure is that for low-mass systems, the circular SNR
curve seems to overlap that of eccentric binaries. One can understand this by
noting that low-mass systems merge in the high-frequency band of the detector,
since the merger frequency is inversely proportional to the total mass. In
such cases, the binary circularizes before merger. Suppose that the binary's
eccentricity reduces to $e_{0} \leq 10^{-2}$ at some ``circularization
frequency'' $F_{c}$. One can then compare the number of cycles the binary
spends in $\{f_{\rm low},F_{c}\}$ relative to the number of cycles spent in
$\{F_{c},F_{\rm high}\}$, to find that the latter is overwhelmingly large for
low-mass systems. Such a fact does not imply that circular waveforms are
sufficient for detection or parameter estimation to extract signals from
eccentric inspirals of low-mass. The SNRs shown here are ``optimal,'' and
thus, a much more careful fitting-factor study is necessary to determine
whether circular templates suffice to extract eccentric binary signals.

For high masses the SNR presents a somewhat oscillatory behavior. These
oscillations seem to scale with the eccentricity, becoming worse for systems
with $e_{0} = 0.4$. Oscillations are expected, since different harmonics could
interfere in the SNR integrand and since the step-function truncation of the
waveforms will introduce oscillations at overtones of the truncation
frequencies.  We discuss these issues in more detail in the next subsection.

\subsection{Accuracy of the Approximation}

An important issue concerns the accuracy of our post-circular approximation.
Our approximation is essentially an expansion for $e_{0} \ll 1$, so it should
break down as we increase the initial eccentricity. On the other hand, if the
condition $e_0 \ll 1$ is verified, a relatively small number of harmonics
should model the waveform accurately enough that we would not lose much in
terms of SNR.  

\begin{figure*}[htp]
\begin{center}
\begin{tabular}{cc}
  \includegraphics[width=8cm,clip=true]{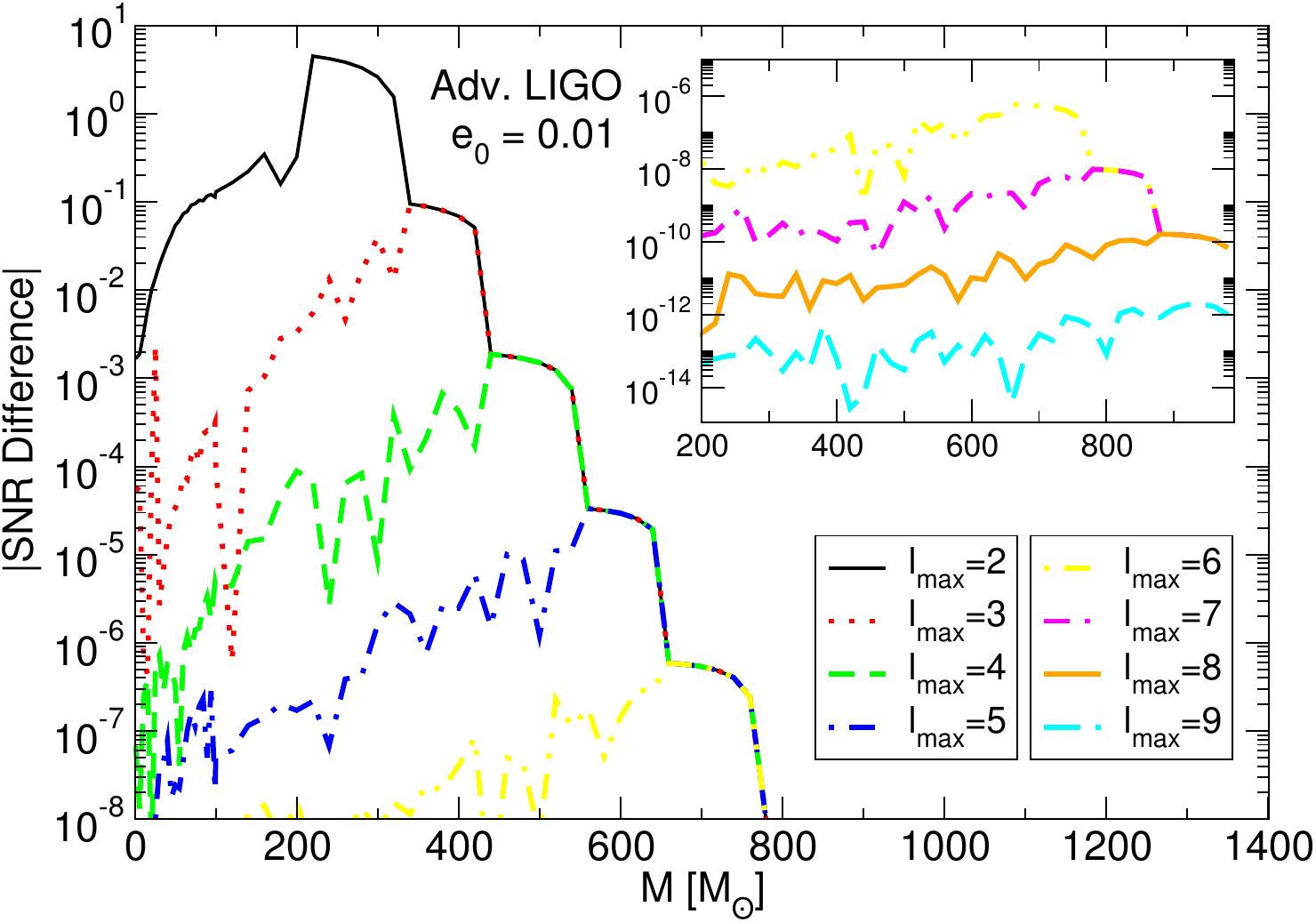}  &
  \includegraphics[width=8cm,clip=true]{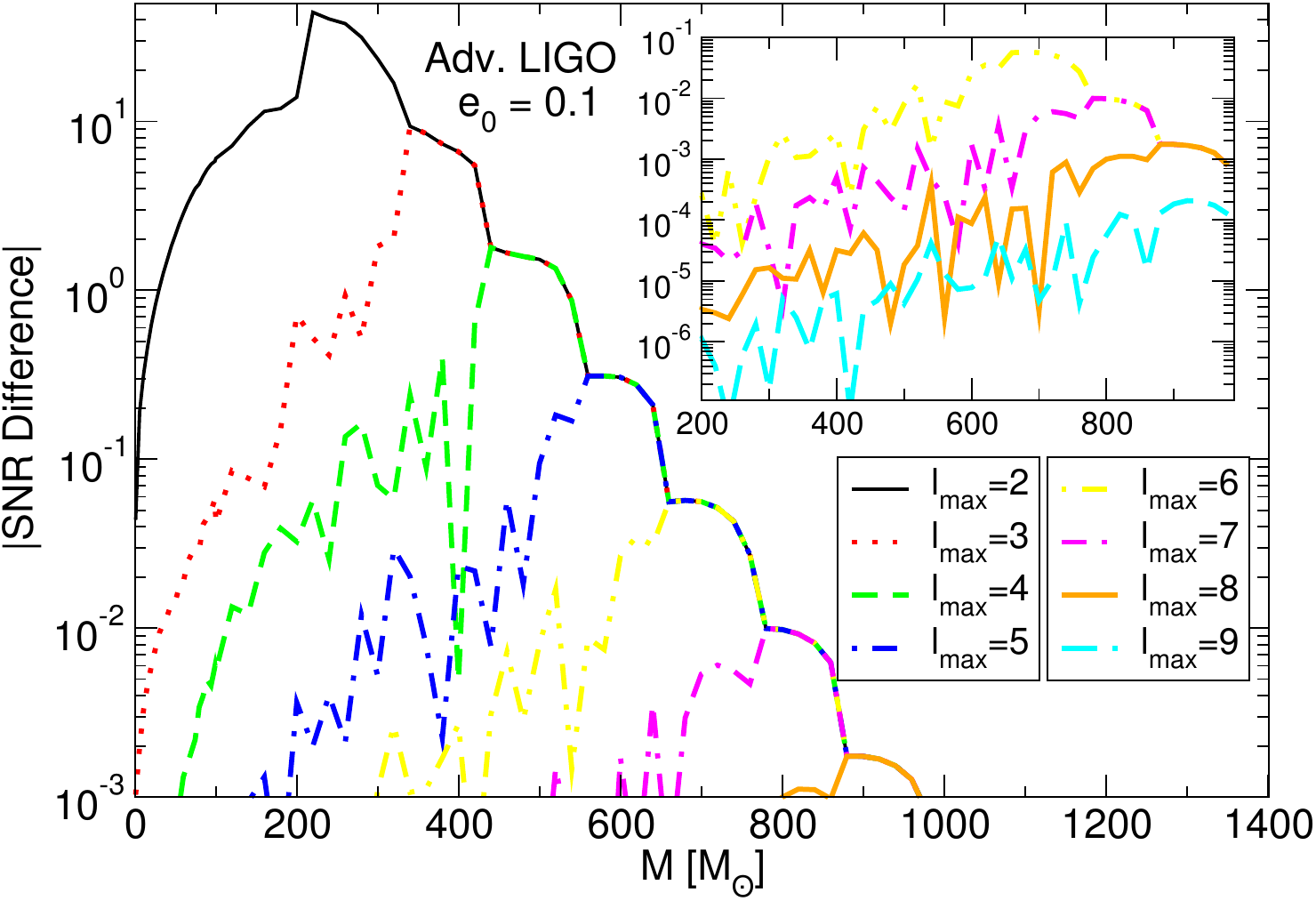} \\
  \includegraphics[width=8cm,clip=true]{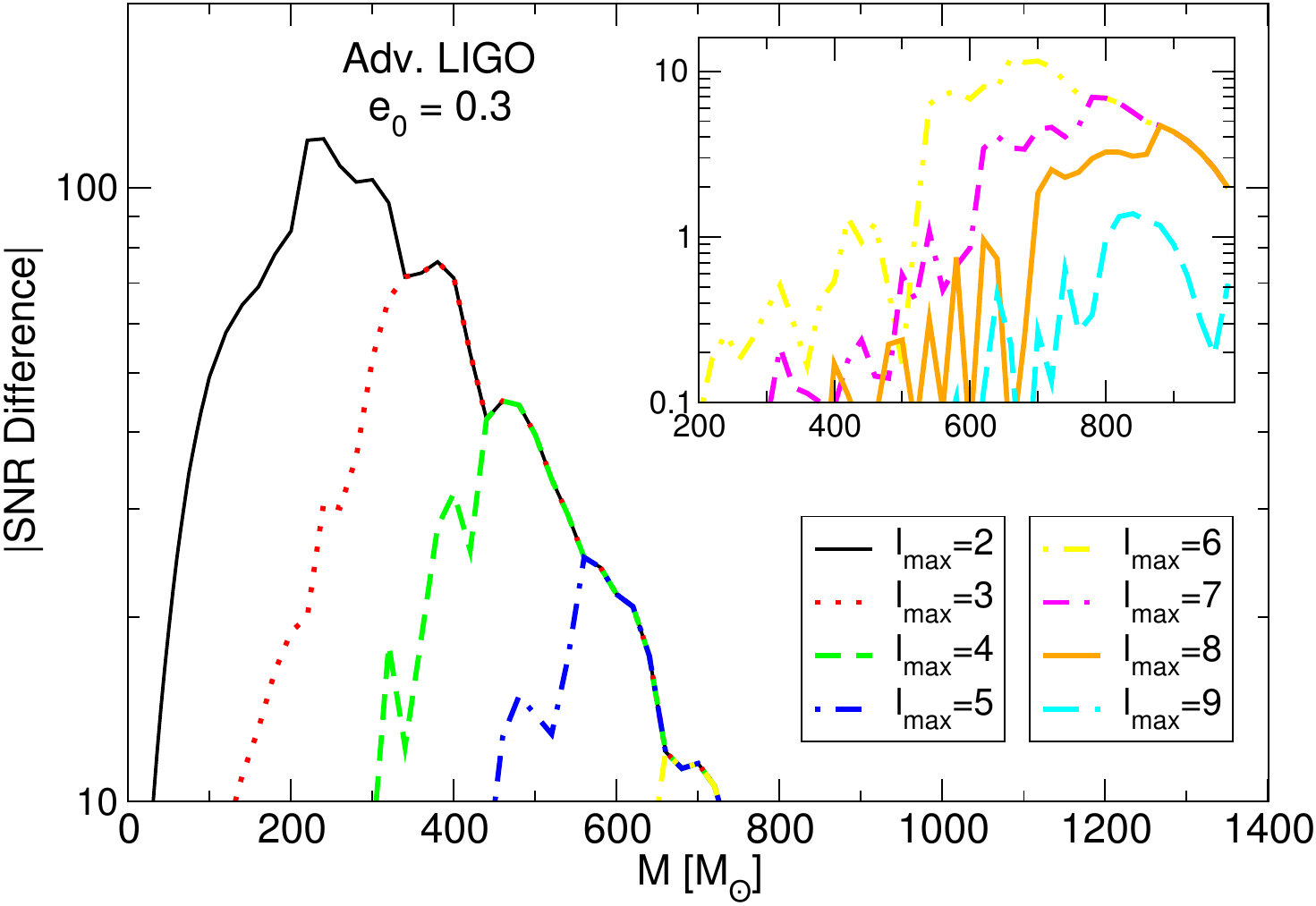} &
  \includegraphics[width=8cm,clip=true]{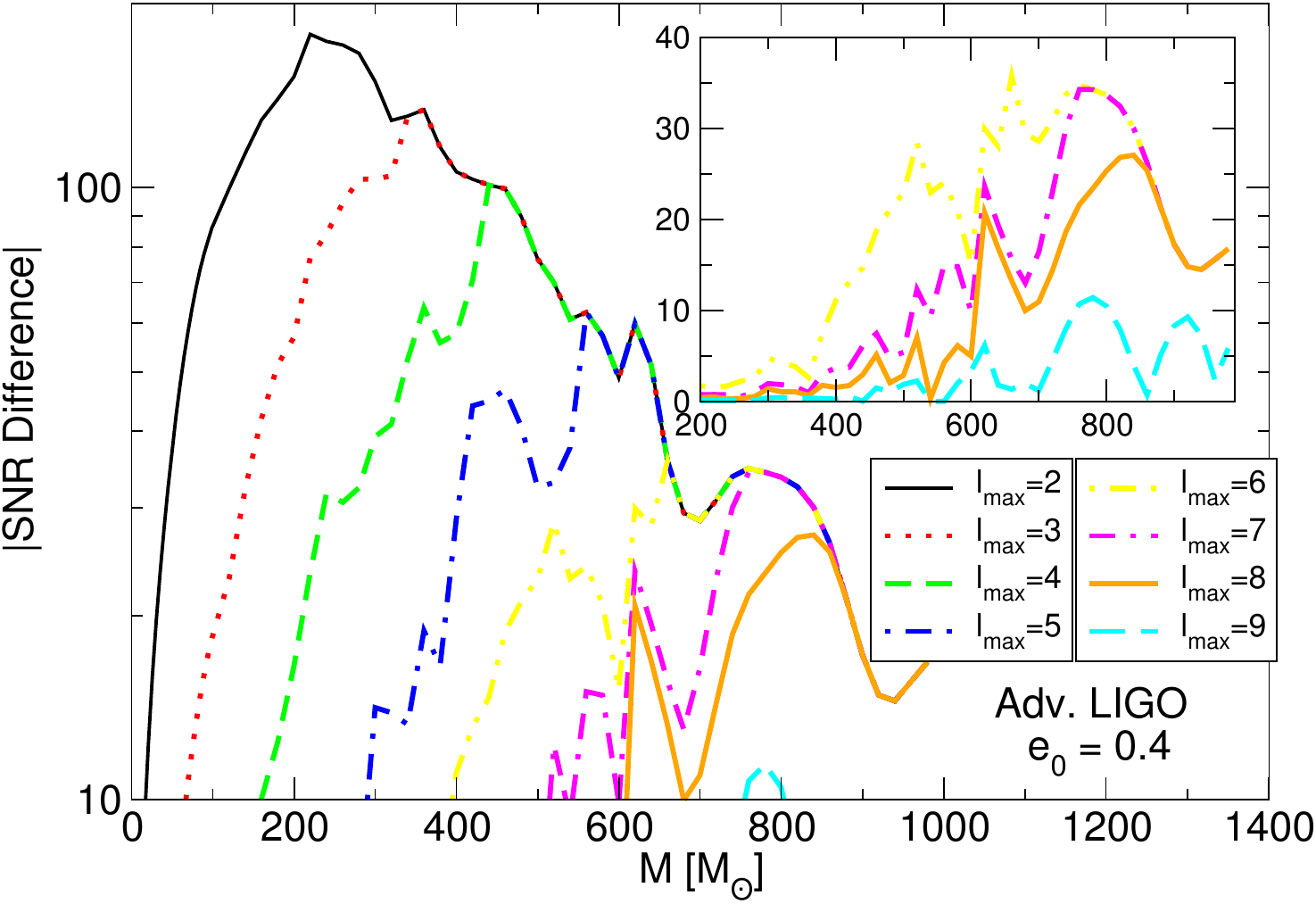}
\end{tabular}
\end{center}
\caption{\label{SNR-diffk} Absolute value of the difference in SNR computed up
  to $\ell_{\rm max}=(2,3,4,5,6,7,8,9)$ and up to $\ell_{\rm max} = 10$. We
  use the AdvLIGO noise curve and an initial eccentricity of $e_{0} = 0.01$,
  $e_{0} = 0.1$, $e_{0} = 0.3$ and $e_{0} = 0.4$ (top to bottom). The insets zoom 
  on the $y$-axis in the region between $200$ and $800$ solar masses.}
\end{figure*}

In order to explore this issue, in Fig.~\ref{SNR-diffk} we plot the absolute
value $\delta \rho(\ell_{\rm max},10)\equiv |\rho(\ell_{\rm max})-\rho(10)|$,
where $\rho(\ell_{\rm max})$ is the SNR computed by keeping $\ell_{\rm max}$
terms in the harmonic sum of Eq.~\eqref{fft}.  In the top two panels we
consider systems with moderate initial eccentricity ($e_{0} = 0.01$ and $e_{0}
= 0.1$), which are probably most relevant for several classes of astrophysical
GW sources.  When $e_{0} = 0.01$, the deviation in SNR relative to the
highest-order terms we computed ($\ell_{\rm max} = 10$) is at most of
${\cal{O}}(1)$ or of ${\cal{O}}(10^{-1})$ when one uses $\ell_{\rm max} = 2$
and $\ell_{\rm max} = 3$, respectively.  On the other hand, for the $e_{0} =
0.1$ case, a comparable accuracy in SNR requires $\ell_{\rm max} \geq 4$ and
$\ell_{\rm max} \geq 5$, respectively. It should not be surprising that a
smaller number of harmonics is required for systems with low eccentricity. Our
analysis suggests that summing up to $\ell_{\rm max} \simeq 4$ should be enough
for systems with $e_{0} \leq 0.1$, while for systems with $0.1 < e_{0} \leq
0.3$ one needs $\ell_{\rm max} \geq 8$.

More interesting features emerge for larger values of $e_0$ (bottom
panels in Fig.~\ref{SNR-diffk}), which we list below:
\begin{enumerate}
\item Different harmonics play a critical role in the SNR at different mass
  ranges. When $e_{0} = 0.3$ the SNR difference peaks at approximately
  $(200,400,500,600) M_{\odot}$ for $\ell_{\rm max} = (2,3,4,5)$; for lower
  and larger masses, a smaller number of harmonics is necessary.

\item The number of harmonics necessary to cover the entire mass range is a
  function of the initial eccentricity $e_0$. When $e_{0} = 0.3$, for example,
  harmonics with $\ell >7$ are not needed, because $\delta \rho(6,10) < 7$ in
  the entire mass range. On the other hand, for the $e_{0} =0.4$ case, one
  really needs at least $\ell_{\rm max} = 9$ to obtain errors $\delta
  \rho(9,10) < 10$ in the whole mass range, while for $e_{0} = 0.5$ the
  approximation seems to break down, unless more harmonics are included.

\item The oscillations visible in the plots are not necessarily an artifact of
  the post-circular approximation. Indeed, these oscillations are also present
  in the small eccentricity curves [$e_{0} = (0.1,0.2,0.3)$] of Fig.~\ref{SNR}
  and in the top panel [$e_{0} = (0.01,0.1)$] of Fig.~\ref{SNR-diffk}.  If
  these oscillations were an artifact of the post-circular approximation, they
  would vanish in the small eccentricity limit, but instead, although they
  decrease in magnitude, they are still present.

\end{enumerate}

The fact that different harmonics peak at different masses can be understood
by observing that two competing effects control the SNR difference: the
eccentricity decay and the frequency band over which we perform the
integration.  For small masses, one is integrating over a larger frequency
band, and the binary rapidly circularizes before merging. Less harmonics are
needed in the limit of very small mass, since the binary is essentially
circular before reaching the most sensitive region of the detector.  On the
other hand, for really high masses one is integrating for short times and
essentially capturing only the behavior near the ISCO. In such cases, the SNR
difference is converging to zero, because the SNRs themselves are essentially
vanishing (i.e.~the range of integration asymptotes to zero).

The oscillatory features can be understood by studying the analytic structure
of the waveforms used to compute the SNRs.  For small eccentricities, the
oscillations are probably due to interference between the different harmonics
and to the use of step functions to truncate the SNR at different harmonics.
The waveform contains a sum of ten different oscillatory functions, and when
this sum is multiplied by its complex conjugate, one naturally obtains
interference of the type $\exp\left[ i (\ell - \ell') t\right]$. Moreover, the
step function truncation of the SNR also forces oscillations at overtones of
the ISCO frequency. For example, Figures \ref{SNR} and \ref{SNR-diffk} show
oscillations at $\ell F_{\ISCO}$, with $\ell = \left\{1,10\right\}$. This is
because the leading harmonic has a mass reach corresponding to twice the ISCO
frequency, whereas the $\ell$--th harmonic has a mass reach corresponding to
$\ell$ times the ISCO frequency. Since we use a step function cut-off for
every harmonic [see e.g. Eq.~(\ref{eq:AL-ET})], the resulting SNR will show
the ``bumps'' seen in Fig.~\ref{SNR-diffk}. Of course, although there are
analytic reasons that explain the presence of these oscillations, one cannot
formally exclude the possibility that (for large $e_0$) some of these
oscillations are induced by inaccuracies in the post-circular approximation.

\subsection{Angular dependence}
\label{Sec:angulardep}

\begin{figure*}
[htp]
\begin{center}
\begin{tabular}{c}
\includegraphics[width=8cm,clip=true]{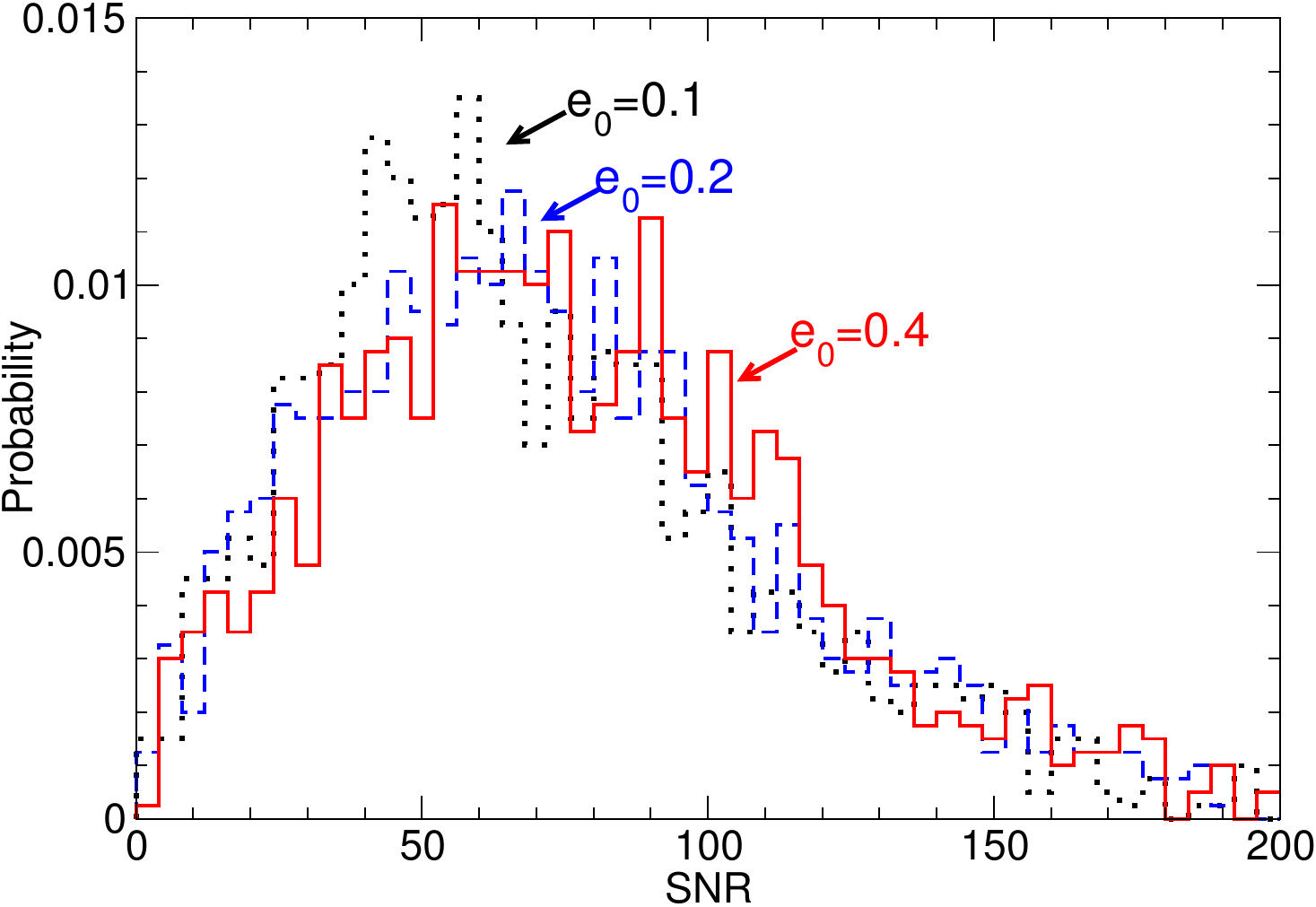}
\includegraphics[width=8cm,clip=true]{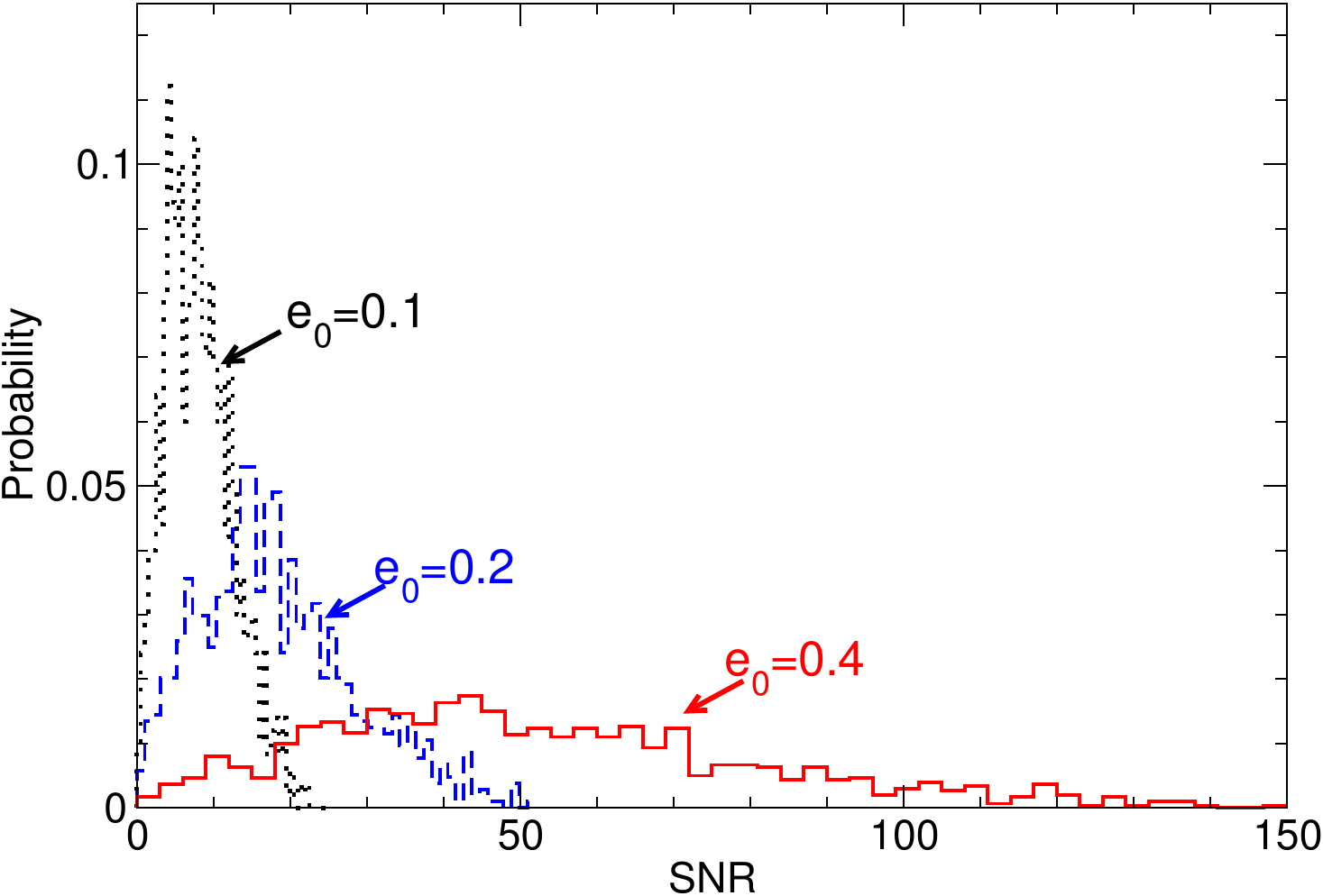}
\end{tabular}
\end{center}
\caption{\label{orientation} Histograms of the AdvLIGO SNR  (see
  text) for a set of random observer orientations. The left (right) panel
  corresponds to a $M=100M_\odot$ ($M=300M_\odot$) system. Systems with
  initial eccentricity $e_{0} = (0.1,0.2,0.4)$ are shown in dotted black,
  dashed blue and solid red, respectively.}
\end{figure*}

In all SNR plots, we have so far assumed that binaries are optimally
oriented. This is a very special configuration, and it is important to
investigate the variations in SNR for non-optimally oriented binaries. Figure
\ref{orientation} shows histograms of the SNR distribution under the
hypothesis of a uniform distribution of the angles involved: the angles
$(\iota,\beta)$ describing the orientation of the source and the three angles
($\theta_S$, $\phi_S$, $\psi_S$) appearing in the antenna pattern functions
\cite{BBW05a}. We binned the data of the 1000 random realizations in bins of
50. The histograms are representive of the number of realizations in each bin.
Since they are normalized, the numbers on the vertical axis do not correspond
to the actual number of realizations in each bin. The comparison of
distributions corresponding to various eccenricities makes sense only if the
histograms are normalized.
In the two panels of Fig.~\ref{orientation} we plot the AdvLIGO SNR
distribution for two representative binaries with total mass 100$M_\odot$
(left panel) and 300 $M_\odot$ (right panel). The SNR of the 100 $M_\odot$
system is dominated mostly by the leading harmonic, whereas the 300 $M_\odot$
system has significant contribution from higher harmonics (see the top panel
of Fig.~\ref{SNR}).  In each panel we show three histograms, corresponding to
initial eccentricities $e_0=0.1$ (dotted black), $0.2$ (dashed blue) and $0.4$
(solid red).

The effect of eccentricity is to shift the SNR distribution to higher masses.
For the 100$M_\odot$ system, although the SNR distribution visibly shifts to
the right, the effect is rather mild for the initial eccentricities considered
in the plot.  In this case, the distributions look more or less similar,
except for a longer tail at large SNRs for large initial eccentricities. For
the 300$M_\odot$ system, the shift in the distribution is much more pronounced. While
the distribution is narrowly peaked at low eccentricities, it becomes much
wider for larger eccentricities. This widening of the distribution might be
because, for large eccentricities, the waveforms are more sensitive to terms
proportional to $c_i=\cos \iota$ (see the expressions of $C_{+}^{(n)}$ and
$S_{+}^{(n)}$ in Appendix \ref{app:ccoeffs}). These amplitude corrections,
being proportional to the eccentricity, are suppressed for small values of
$e_0$.

The SNR histograms show that, although the SNR values for any particular observation may
deviate considerably from the optimal values we have quoted, the general
trends in the SNR induced by eccentricity (mass reach increase, accessible volume increase) 
will not be modified. That is, regardless of the location of the source in the sky, eccentricity has the net effect of
increasing the SNR or mass reach of any given signal.

%

\section{PN corrections} 
\label{Sec:PN}

Although this paper has only considered Newtonian waveforms to exemplify the
post-circular construction of the SPA Fourier transform of eccentric
waveforms, it is instructive to discuss how to extend these results to higher
PN order.  In this section we shall mainly follow the conventions of Gopakumar
and Iyer~\cite{Gopakumar:2001dy}, where a ``2PN term'' is one of
${\cal{O}}(\dot{r}/c)^4$ smaller than the leading-order term, i.e.~we employ 
a {\emph{relative}} order counting scheme.

When eccentric orbits are studied beyond $1$PN order, one discovers that the
Keplerian parameterization must be corrected. In particular,
Eqs.~(\ref{radius-eq}) and~(\ref{radius-eq3}) are enhanced by new
${\cal{O}}(\dot{r}/c)^2$ corrections, while Eq.~(\ref{radius-eq2}) is enhanced
by corrections of ${\cal{O}}(\dot{r}/c)^4$. Moreover, if we want to keep a
Keplerian-inspired parameterization we must introduce {\it three} distinct
eccentricity parameters: $e_r$ measures radial oscillations, $e_{\phi}$
measures azimuthal oscillations and $e_t$ measures the frequency
eccentricity. These eccentricity parameters can be written as functions of
$e_t$ in a PN expansion, but this choice is somewhat arbitrary (see
e.g.~\cite{JenaEccentric07} for comparisons of the different definitions with
numerical relativity simulations of eccentric mergers).

The waveforms then acquire two sets of modifications: amplitude and phase
corrections.  The amplitude corrections, as expected, take the form
\be
\label{PN-WFs}
h_{+,\times} \approx \frac{m \eta}{R} \zeta^{2/3} \left[ H_{+,\times}^{(0)}
  + \zeta^{1/2} H_{+,\times}^{(1/2)} + \zeta H_{+,\times}^{(1)} +
  \ldots \right],
\ee
where $\zeta \equiv m N$ is a natural PN expansion parameter of
${\cal{O}}(\dot{r}/c)^2$, $\eta = \mu m$ is the symmetric mass ratio and
$H_{+,\times}^{(n)}$ are products of functions of the eccentricities and
harmonic trigonometric functions. The phase corrections can be split into
\be
\label{corrected-phase}
\phi - \phi_0 = \lambda(l) + W(l),
\ee
where $\lambda(l)$ is a $2 \pi K$-periodic function of the mean anomaly, while
$W(l)$ is periodic in $l$, and thus, $2 \pi$-periodic.  The quantity $K$
measures the advance of the periastron per orbital
revolution~\cite{ABIQ07tail}, while the effect of $W(l)$ is to modulate the
amplitude via nutation.

The leading order correction to the phasing of GWs is due to pericenter
precession. This effect is embodied in the function $\lambda(l)$, which can be
written as
\be
\label{1PNlambda}
\lambda(l) = K l = \left[1 + k_p(e_t)\right] l,
\ee
where we have defined $K = 1 + k_p(e_t)$. To this order, the mean anomaly
continues to be given by Eq.~\eqref{radius-eq2} with $e \to e_{t}$, while the
precession correction $k_p(e_t)$ is given by
\be
k_p(e_t) = \frac{3 \zeta^{2/3}}{1 - e_t^2} + {\cal{O}}(\dot{r}/c)^4.
\ee
Since $N$ appears here as a $1$PN order correction,
we can take its Newtonian value in this equation, namely $N = 2 \pi f$, such
that $\zeta = 2 \pi M F$. Thus, the precession correction becomes
\be
k_p(e_t) = \frac{3 \left(2 \pi M F \right)^{2/3}}{1 - e_t^2} +
{\cal{O}}(\dot{r}/c)^4.
\ee
We can expand this function for small eccentricities $e_t \ll 1$ to find
\beq
\lambda(l) \sim l \left[ 1 + 3 \left(2 \pi M F \right)^{2/3} \left(1 +
    e_t^2 + e_t^4 + e_t^6 + e_t^8 \right) \right].
\eeq

The second correction to the phase is given by the nutation function $W$,
defined by
\be
\label{W-def}
W = \left( v - u + e_t \sin{u}\right) \left[1 + k_p(e_t) \right] +
{\cal{O}}(\dot{r}/c)^4, 
\ee
where the true anomaly is given by
\be
\label{corrected-v}
v = 2 \tan^{-1} \left[ \left(\frac{1 + e_{\phi}}{1 -
      e_{\phi}}\right)^{1/2} \tan\left(\frac{u}{2}\right)\right].
\ee
Note that the true anomaly depends on $e_{\phi}$ and not $e_t$, but these
quantities are related via
\be
\label{ephi}
e_{\phi} = e_t \left[1 + \zeta^{2/3} \left(4 - \eta\right) +
  {\cal{O}}(\dot{r}/c)^4\right]. 
\ee
In Sec.~\ref{Sec:kepler} we already discussed how to solve for $u$ as a
function of $l$ in terms of a series of Bessel functions. In particular,
one can show that
\be
\label{u-Bessel}
u = l + \sum_{s=1}^{\infty} \left(\frac{2}{s}\right) J_s\left(s
  e\right) \sin{\left(s l\right)}.
\ee
Equations \eqref{ephi} and (\ref{u-Bessel}) can be substituted in
Eq.~(\ref{corrected-v}) to find $v$ as a function of $l$. Then
Eq.~\eqref{W-def} yields $W$ as a function of $l$. Expanding in $e_t \ll 1$
one finds
\beq
W(l) &\sim& e_t \left[ - \left( -10+\eta \right) {\zeta}^{2/3}+2 \right] \sin
 \left( l \right)
\nonumber \\
&+& {e_t}^{2} \left[ -1/2\, \left( 4\,\eta-31 \right) {\zeta}^{2/3}+5/2
 \right] \cos \left( l \right) \sin \left( l \right) 
\nonumber \\
&+& {e_t}^{3} \left[  \left( -1/6\, \left( -186+27\,\eta \right) {\zeta}^{2/
3}+13/3 \right)  \cos^2 \left( l \right)  
\right. 
\nonumber \\ 
&-& \left. 
1/6\, \left( 12-6\,\eta \right) {\zeta}^{2/3}-4/3 \right] \sin \left(
l \right) + \ldots
\label{W-eq}
\eeq
This function, however, is part of the phase, so it enters the waveform as the
argument of trigonometric functions. Note that $W(l)$ is linear in $e_t$, and
thus, when the $\cos(\phi)$ or $\sin(\phi)$ are expanded in $e_t \ll 1$,
$W(l)$ introduces higher harmonics into the waveforms.

We see then that to $1$PN order, it suffices to consider the pericenter
precession correction through $\lambda$. The corrections produced by $W(l)$
are automatically accounted for in the Bessel expansion. In essence, this is
because Eqs.~\eqref{radius-eq} and~\eqref{radius-eq2} are not modified to this
order.  The $1$PN-corrected waveforms are then (schematically)
\beq
h(t) &=& {\cal{A}} \sum_{\ell=1}^{10} \alpha_\ell \cos\left\{\ell \; l \left[1 + 3
    \left(2 \pi M F\right)^{2/3} \left(1 + t_2 e_t^2 
\right. \right. \right.
\nonumber \\
&+& \left. \left. \left.  t_4 e_t^4  + t_6 e_t^6  \right) \right] + \phi_{\ell}
\right\},
\label{h-t-PN}
\eeq
where the $t_k$'s are constants.  If we were to consider $2$PN corrections to
the waveforms, then the formalism outlined here would have to be extended and
$W(l)$ would contribute by introducing new corrections not accounted for in
the Bessel expansion.

The structure of the 1PN time-domain waveform in Eq.~\eqref{h-t-PN} is
different from that obtained in
Refs.~\cite{1994MNRAS.266...16M,1995MNRAS.274..115M}, in that the above
equation does not lead to periastron-precession side-bands in the GW
spectrum. In Refs.~\cite{1994MNRAS.266...16M,1995MNRAS.274..115M} such
side-bands arise due to the assumption that periastron precession leads to a
constant $\dot{\gamma} \propto k_{p}$.  This assumption breaks down on long
time-scales as periastron precession is not constant, an effect one can justly
treat as a $1$ PN contribution. As a result, one loses the artificial
side-band structure in the GW spectrum. The implications of this effect will
be assessed in future work.


PN corrections modify the Fourier transform of the waveform in the SPA. The
phase $\bar\Psi_\ell$ is modified by a factor of $(1 + k_p)$, and we now
obtain
\begin{eqnarray}
\Psi_\ell(F) &=& k \; \lambda[t(f/\ell)] - 2 \pi f t(f/\ell),
\end{eqnarray}
where
\beq
\lambda[t(f/\ell)] &=& \ell \phi_c + \ell \int^{f/\ell} \frac{\dot\lambda'}{\dot{F}}  dF',
\nonumber \\
t(f/\ell) &=& t_c + \int^{f/\ell} \frac{dF'}{\dot{F}'}.
\eeq
The $\dot{\lambda}$ term contains the $(1 + k_{p})$ dependence that we
referred to via Eq.~\eqref{1PNlambda}. All the machinery developed in the
previous section then carries through, with the proper enhancement of the
Newtonian waveform to higher PN order. The net effect of higher PN corrections
in the Fourier transform is to introduce an infinite set of harmonics and PN
corrections to the lower-order (Newtonian) harmonics considered earlier.


While considering higher PN order effects, one can also work with the PN
parameter $(M\omega)^{2/3}$, where $\omega$ is the orbital frequency. As
pointed out in Ref.~\cite{ABIQ07}, this parametrization helps to more easily
recover the circular limits of various elliptic-orbit expressions.
Furthermore, in comparing numerical relativity results to PN expansions,
Ref.~\cite{HinderPNeccentric08} found that waveforms parametrized in terms of
$(M\omega)^{2/3}$ are in better agreement with numerical waveforms. This
deserves more careful study in the future.

\section{Conclusions} 
\label{Sec:conclusions}

We have proposed a new scheme, the post-circular approximation, to construct
``ready-to-use'', analytic Fourier-domain gravitational waveforms produced
by eccentric binary inspirals.  The scheme consists of expanding all
quantities in a power series about zero initial eccentricity.  We find that
the first $10$ terms in the Bessel solution to the Kepler problem suffice to
reproduce the eccentricity evolution to better than $0.1 \%$ for
eccentricities $e < 0.4$.  The resulting waveforms are then rewritten in terms
of ten physical parameters (the reduced mass $\eta$, the initial eccentricity
$e_0$ and frequency $F_0$, the total mass $M$, the luminosity distance $D_L$,
four angles $\iota$, $\beta$, $\theta$, $\phi$ describing the relative
orientation of the source and detector, and a polarization angle $\psi$) and
the orbital frequency, which can be thought of as a function of time.

This scheme allows us to analytically construct the Fourier transform of the
response function through the SPA, where one assumes that the
radiation-reaction time scale is much larger than the orbital time scale.  The
resulting Fourier-domain waveforms contain eccentricity-induced,
higher-harmonic amplitude and phase corrections. By computing the SNR as a
function of total mass and eccentricity we find that the amplitude corrections
increase the mass reach of the detectors by a factor $\simeq 5$ for moderately
eccentric systems, which in turn implies that the source volume accessible to
the detectors would be increased by almost two orders of magnitude.

The results presented here cannot be used directly in realistic data analysis
pipelines because PN corrections to the amplitude and phase have not been
included. Instead, the present paper was concerned with proposing a method to
construct ``ready-to-use'', analytic expressions for the Fourier transform of
the response function, which was exemplified through Newtonian-accurate
expressions. Future research should include such PN corrections.

Another interesting research direction is the study of the effect of
eccentricity in parameter estimation. Eccentricity adds more complexity and
information to the waveforms that could break parameter degeneracies, thus
possibly leading to better accuracy in parameter estimation. On the other
hand, the inclusion of eccentricity-induced corrections to the GW phase could
mimic certain high-order PN phase corrections, which would then create new
degeneracies. A more detailed parameter estimation study is needed to assess
whether GW measurements will benefit or not from the inclusion of
eccentricity.

While finalizing the draft we learned that two different groups are
investigating frequency-domain gravitational waveforms for eccentric
binaries~\cite{favata,kocsis}. It would be interesting to compare their
approach with ours and with the time-domain waveforms of~\cite{DGI04}.

\section*{Acknowledgements}

We thank Eric Poisson, Sai Iyer, David Spergel and Frans Pretorius for very
useful discussions.
KGA and CMW were supported in part by the National Science Foundation, Grant
No.\ PHY 06-52448, the National Aeronautics and Space Administration, Grant
No.\ NNG-06GI60G, and the Centre National de la Recherche Scientifique,
Programme Internationale de la Coop\'eration Scientifique (CNRS-PICS), Grant
No. 4396.
NY acknowledges support from the NSF grant PHY-0745779.


\appendix

\section{\label{app:LISA}LISA eccentric binaries}

In this Appendix we briefly review some literature on scenarios leading to
non-eccentric binary inspirals in the LISA band. We consider in turn stellar
mass binaries, extreme- and intermediate-mass ratio inspirals (EMRIs/IMRIs,
respectively), and the coalescence of massive BHs.

\subsection{Stellar mass binaries} 
Many stellar mass binaries involving neutron stars are expected to be
eccentric in the LISA band (see e.g.~\cite{Benacquista:2002kf}). It is also
well known that LISA should provide a large observational sample of
interacting white-dwarf binaries, whose evolution is driven by radiation
reaction, tides and mass transfer~\cite{Stroeer:2005cv}.  It was recently
realized that eccentric double white dwarfs formed in globular clusters would
be detectable by LISA out to the Large Magellanic Cloud
\cite{Willems:2007nq}. In these binaries, the periastron precession has
contributions due to general relativity, but also to tidal and rotational
distortions. Tides and stellar rotation should dominate at frequencies above a
few mHz. The Fisher-matrix analysis of~\cite{Willems:2007xe} pointed out the
interesting possibility to study white dwarf structure with LISA. However
their analysis neglected the contribution of radiation reaction effects, that
should be relevant for $f\gtrsim 0.5$~mHz. We expect our post-circular
formalism to be useful in this context, since radiation reaction in these
eccentric binaries should be well modeled by the quadrupole approximation.

\subsection{Extreme and intermediate mass ratio inspirals} 
Formation scenarios for EMRI and IMRIs, involving a SMBH and either a compact
stellar mass object or an intermediate-mass BH (IMBH), are reviewed in
Ref.~\cite{AmaroSeoane:2007aw}. If an EMRI is formed following a tidal binary
separation event, the compact star is deposited on an orbit with semi-major
axis $\simeq 10^2-10^3$~AU and $e\simeq 0.9-0.99$, and the orbit should
circularize by the time it enters the LISA band. However, typical EMRIs are
expected to form by scattering of the compact object into nearly radial orbits
followed by inspiral due to dissipation, and in particular due to GW-emission.
Hopman and Alexander \cite{Hopman:2005vr} showed that the eccentricity
distribution of EMRIs is skewed to high-$e$ values, with a peak at $e\simeq
0.7$, at an orbital period of $\simeq 10^4$~s.
The dynamical evolution of IMBH binaries formed in dense stellar clusters,
using a combination of $N$-body simulations and three-body relativistic
scattering experiments, shows that the eccentricity of these systems in the
LISA band can be as large as $\simeq 0.2-0.3$ \cite{AmaroSeoane:2009yr}. The
post-circular approximation developed here could be applied to these systems
once PN corrections are taken into account.

\subsection{Massive black hole coalescence} 
The eccentricity of SMBH binaries has been the subject of some
debate. Gravitational radiation reaction alone is not sufficient to produce
mergers between massive BHs, which probably require dynamical interactions.
Analytic calculations and $N$-body simulations show that, in purely
collisionless spherical backgrounds, the expected equilibrium distribution of
eccentricities is skewed towards high $e\simeq 0.6-0.7$, and that dynamical
friction does not play a major role in modifying such a distribution (see
Ref.~\cite{ColpiMayer}, in particular Fig.~5). The actual eccentricity of a
merger event is therefore determined by the competition between dynamical
wandering and GW-induced circularization. Reference~\cite{AmaroSeoane:2007aw}
presents arguments supporting circularization of most binaries by the time
they enter the LISA band.  However, several mechanisms producing non-zero
eccentricity have been proposed in the past (see e.g. Section 2 of
Ref.~\cite{Berti:2006ew}).

Recent smoothed-particle hydrodynamics simulations follow the dynamics of two
BHs orbiting in massive, rotationally supported circumnuclear discs
\cite{Dotti:2005kq,Dotti:2006ef}. The rotation of the disc circularizes the
orbit if the pair {\it corotates} with the disc. Circularization is efficient
until the BHs bind in a binary, though in the latest stages of the simulations
(when the separation is of the order of a few parsecs) a residual eccentricity
$e\gtrsim 0.1$ is still present. Circularization possibly reduces the
gravitational radiation merging time scale so much that the binary stalls, and
no coalescence results. For corotating discs, the numerical resolution of the
simulations is not sufficient to compute the residual eccentricity when the
BHs are close enough that gravitational radiation takes over. Moreover, if the
orbit of the pair is {\it counterrotating} the initial eccentricity does not
decrease, and BHs may enter the GW-dominated phase with high eccentricity.

Collisional processes (such as three-body encounters with background stars)
may become important at BH separations $\lesssim 6$~pc, possibly leading to
an increase in eccentricity balancing the circularization driven by the
large-scale action of the gaseous and/or stellar disc. Several investigations
show that eccentricity evolution may still occur in later stages of the
binary's life, because of close encounters with single stars
\cite{Berczik:2006tz} and/or gas-dynamical processes \cite{Armitage:2005xq}.
In particular, the gravitational interaction of the binary with a surrounding
gas disc is likely to excite BH binaries to eccentricities $e\gtrsim 0.1$. The
transition between disc-driven and gravitational wave-driven inspiral can
occur at small enough radii that a small but significant eccentricity
survives, with typical values $e\simeq 0.02$ (and a lower limit of $e\simeq
0.01$) one year prior to merger (cf. Fig.~5 of \cite{Armitage:2005xq}). If the
binary has an extreme mass ratio $q\lesssim 0.02$ the residual eccentricity
can be considerably larger ($e\gtrsim 0.1$). Recent simulations by Cuadra {\it
  et al.}  \cite{Cuadra:2008xn} investigate the evolution of the orbital
parameters of binaries embedded within geometrically thin gas disks. For
binary masses $10^5M_\odot\lesssim M \lesssim 10^8M_\odot$, they find that
orbital decay due to gas disks may dominate the binary dynamics for
separations below $a\simeq 10^{-1}-0.1$~pc, and that in the process the
eccentricity grows at a rate $de/dt\simeq 1.5\times 10^{-4}\omega_{\rm orb}$,
where $\omega_{\rm orb}$ is the orbital frequency. Saturation of the
eccentricity growth is not observed up to values $e\gtrsim 0.35$, so the
binary may have significant eccentricity by the time gravitational radiation
takes over.

Stellar dynamical hardening might also leave the binary with non-zero
eccentricity. Early studies suggested that any such eccentricity would be
small \cite{Quinlan:1996vp,Quinlan:1997qe,Berczik:2006tz} (but see
\cite{Matsubayashi:2004bd} and \cite{Aarseth:2002ie} for examples of
eccentricity growth in $N$-body simulations). More recent $N$-body simulations
combined with a Fokker-Planck model \cite{Merritt:2007db} find that
perturbations of the (initially circular) binary orbit from passing stars
produce significant eccentricity around or even before the time when the
binary becomes hard. The averaged eccentricity growth is maximum for
equal-mass binaries with $e\approx 0.75$ and falls to zero at $e=0$ and $e=1$.
It is hard to estimate the final eccentricity, which strongly depends on
noise-induced changes in $e$ at early times, and would presumably be much
smaller than the simulations suggest in the large-$N$ regime of real galaxies.

Berentzen {\it et al.} \cite{Berentzen:2008yw} present simulations following
the SMBH evolution in rotating galactic nuclei from kpc separations down to
coalescence, including post-Newtonian (PN) corrections to the binary equations
of motion. They find that the orbital eccentricities remain large (between 0.4
and 0.99, with typical values around $e\simeq 0.9$) until shortly before
coalescence, and that higher harmonics of the eccentric signal are detectable
by LISA with large SNR. Most of these binaries have sizable eccentricities (up
to $\simeq 0.2$) by the time they reach a separation $\simeq 10^2$ Schwarzschild
radii, which roughly corresponds to a binary of mass $M\approx 2\times
10^6~M_\odot$ entering the LISA band (see e.g.~Figure 8
of~\cite{Berentzen:2008yw}).
The study by Sesana {\it et al.}
\cite{Sesana:2006xw,Sesana:2006ne,Sesana:2007vr} confirms that binaries with
mass ratio $q=M_2/M_1\lesssim 0.1$ and/or eccentricity $e\gtrsim 0.3$ can
shrink to the GW-dominated regime within a Hubble time (see in particular
Figure 7 of \cite{Sesana:2006ne}; Section 4.1 and Figure 10 of
\cite{Sesana:2007vr}). 
Last but not least, an interesting scenario producing highly eccentric mergers
that could be observed by LISA involves close {\em triple} SMBH encounters
\cite{Hoffman:2006iq,Iwasawa:2008hn}.

\section{\label{app:ccoeffs}Higher-order coefficients}

In this Appendix we list some of the higher-order coefficients appearing in
the expansion (\ref{expansion}).

\begin{widetext}
\beq
\label{C+}
C_+^{(2)} &=& s_i^2 \left(e^2 - \frac{1}{3} e^4 + \frac{1}{24} e^6 -
  \frac{1}{360} e^8 \right) + \left(1 + c_i^2\right) c_{2 \beta}
\left(2 - 5 e^2 + \frac{11}{4} e^4 - \frac{179}{360} e^6 +
  \frac{7}{160} e^8 \right),
\\
C_+^{(3)} &=& s_i^2 \left(\frac{9}{8} e^3 - \frac{81}{128} e^5 +
  \frac{729}{5120} e^7 \right) + \left(1 + c_i^2\right) c_{2 \beta}
\left(\frac{9}{2} e - \frac{171}{16} e^3 + \frac{9477}{1280} e^5 -
  \frac{2187}{1024} e^7 \right),
\\
C_+^{(4)} &=& s_i^2 \left(\frac{4}{3} e^4 - \frac{16}{15} e^6 +
  \frac{16}{45} e^8 \right) + \left(1 + c_i^2\right) c_{2 \beta}
\left(8 e^2 - 20 e^4 + \frac{752}{45} e^6 - \frac{688}{105} e^8
\right),
\\
C_+^{(5)} &=& s_i^2 \left(\frac{625}{384} e^5 - \frac{15625}{9216} e^7
\right) + \left(1 + c_i^2\right) c_{2 \beta}
\left(\frac{625}{48} e^3 - \frac{26875}{768} e^5 + \frac{734375}{21504} e^7
\right),
\\
C_+^{(6)} &=& s_i^2 \left(\frac{81}{40} e^6 - \frac{729}{280} e^8
\right) + \left(1 + c_i^2\right) c_{2 \beta}
\left(\frac{81}{4} e^4 - \frac{2349}{40} e^6 + \frac{146529}{2240} e^8
\right),
\\
C_+^{(7)} &=& s_i^2 \frac{117649}{46080} e^7 + \left(1 + c_i^2\right) c_{2 \beta}
\left(\frac{117649}{3840} e^5 - \frac{588245}{6144} e^7 \right),
\\
C_+^{(8)} &=& s_i^2 \frac{1024}{315} e^8 + \left(1 + c_i^2\right) c_{2 \beta}
\left(\frac{2048}{45} e^6 - \frac{48128}{315} e^8 \right),
\\
C_+^{(9)} &=& \left(1 + c_i^2\right) c_{2 \beta} \frac{4782969}{71680}
e^7,
\\
C_+^{(10)} &=& \left(1 + c_i^2\right) c_{2 \beta} \frac{390625}{4032} e^8,
\eeq

\beq
S_+^{(2)} &=& s_{2 \beta} \left(1 + c_i^2\right) \left(2 - 5 e^2 + 3 e^4
- \frac{73}{180} e^6 + \frac{299}{2880} e^8 \right),
\\
S_+^{(3)} &=& s_{2 \beta} \left(1 + c_i^2\right) \left(\frac{9}{2} e
 - \frac{171}{16} e^3 + \frac{9783}{1280} e^5 - \frac{531}{256} e^7
\right), 
\\
S_+^{(4)} &=& s_{2 \beta} \left(1 + c_i^2\right) \left(8 e^2 - 20 e^4
  + \frac{763}{45} e^6 - \frac{4111}{630} e^8 \right), 
\\
S_+^{(5)} &=& s_{2 \beta} \left(1 + c_i^2\right) \left(\frac{625}{48}
  e^3 - \frac{26875}{768} e^5 + \frac{23125}{672} e^7 \right),
\\
S_+^{(6)} &=& s_{2 \beta} \left(1 + c_i^2\right) \left(\frac{81}{4}
  e^4 - \frac{2349}{40} e^6 + \frac{147177}{2240} e^8 \right),
\\
S_+^{(7)} &=& s_{2 \beta} \left(1 + c_i^2\right)
\left(\frac{117649}{3840} e^5 - \frac{588245}{6144} e^7\right),
\\
S_+^{(8)} &=& s_{2 \beta} \left(1 + c_i^2\right)
\left(\frac{2048}{45} e^6 - \frac{48128}{315} e^8\right),
\\
S_+^{(9)} &=& s_{2 \beta} \left(1 + c_i^2\right)
\frac{4782969}{71680} e^7,
\\
S_+^{(10)} &=& s_{2 \beta} \left(1 + c_i^2\right)
\frac{390625}{4032} e^8.
\label{S+}
\eeq

\beq
C_{\times}^{(2)} &=& s_{2 \beta} c_i \left(-4 + 10 e^2 - \frac{11}{2}
  e^4 + \frac{179}{180} e^6 - \frac{7}{80} e^8 \right),
\\
C_{\times}^{(3)} &=& s_{2 \beta} c_i \left(-9 e + \frac{171}{8} e^3 -
  \frac{9477}{640} e^5 + \frac{2187}{512} e^7 \right),
\\
C_{\times}^{(4)} &=& s_{2 \beta} c_i \left(- 16 e^2 + 40 e^4 -
  \frac{1504}{45} e^6 + \frac{1376}{105} e^8 \right),
\\
C_{\times}^{(5)} &=& s_{2 \beta} c_i \left(-\frac{625}{24} e^3 +
  \frac{26875}{384} e^5 - \frac{734375}{10752} e^7 \right),
\\
C_{\times}^{(6)} &=& s_{2 \beta} c_i \left(-\frac{81}{2} e^4 +
  \frac{2349}{20} e^6 - \frac{146529}{1120} e^8 \right),
\\
C_{\times}^{(7)} &=& s_{2 \beta} c_i \left(-\frac{117649}{1920} e^5 +
  \frac{588245}{3072} e^7 \right),
\\
C_{\times}^{(8)} &=& s_{2 \beta} c_i \left(-\frac{4096}{45} e^6 +
  \frac{96256}{315} e^8\right),
\\
C_{\times}^{(9)} &=& - s_{2 \beta} c_i \frac{4782969}{35840} e^7,
\\
C_{\times}^{(10)} &=& - s_{2 \beta} c_i \frac{390625}{2016} e^8,
\label{Cx}
\eeq

\beq
S_{\times}^{(2)} &=& c_{2 \beta} c_i \left(4 - 10 e^2 + 6 e^4 -
  \frac{73}{90} e^6 + \frac{299}{1440} e^8 \right),
\\
S_{\times}^{(3)} &=& c_{2 \beta} c_i \left(9 e - \frac{171}{8} e^3 +
  \frac{9783}{640} e^5 - \frac{531}{128} e^7\right),
\\
S_{\times}^{(4)} &=& c_{2 \beta} c_i \left(16 e^2 - 40 e^4 +
  \frac{1526}{45} e^6 - \frac{4111}{315} e^8\right),
\\
S_{\times}^{(5)} &=& c_{2 \beta} c_i \left(\frac{625}{24} e^3 -
  \frac{26875}{384} e^5+ \frac{23125}{336} e^7\right),
\eeq
\beq
S_{\times}^{(6)} &=& c_{2 \beta} c_i \left(\frac{81}{2} e^4 -
  \frac{2349}{20} e^6 + \frac{147177}{1120} e^8\right),
\\
S_{\times}^{(7)} &=& c_{2 \beta} c_i \left(\frac{117649}{1920} e^5 -
  \frac{588245}{3072} e^7\right),
\\
S_{\times}^{(8)} &=& c_{2 \beta} c_i \left(\frac{4096}{45} e^6 -
  \frac{96256}{315} e^8\right),
\\
S_{\times}^{(9)} &=& c_{2 \beta} c_i \frac{4782969}{35840} e^7,
\\
S_{\times}^{(10)} &=& c_{2 \beta} c_i \frac{390625}{2016} e^8.
\label{Sx}
\eeq
\end{widetext}

\section{\label{app:xik}The $\xi_k$ coefficients}

The coefficients defined in Eq.~(\ref{def:xik}) are given by the following
expressions, when one fixes $\beta = \iota = 0$:
\begin{widetext}
\beq
\xi_{1} &=& -\left[\left( - \frac{19496441}{368640} F_+ -  \frac{20671709}{368640} i F_{\times} \right) e^7 +
\left(\frac{111701}{4608} F_+ + \frac{119365}{4608} i F_{\times} \right) e^5 + \left(-\frac{535}{48} F_+ - \frac{563}{48} i F_{\times}
\right) e^3 \right.
\nonumber \\
&+& \left.3 \left(F_+ + i F_{\times}\right) e\right],
\nonumber \\
\xi_{2} &=&
 \left(\frac{17653698319}{53084160} i F_{\times} +  \frac{17504123791}{53084160} F_+ \right) e^8 +
\left(- \frac{40020301}{276480} i F_{\times} - \frac{39618829}{276480}  F_+ \right) e^6 
\nonumber \\
&+& 
 \left(\frac{8267}{128} F_+ + \frac{8331}{128} i
  F_{\times} \right) e^4 +
\frac{277}{12} \left( - F_+ - i F_{\times}
  \right) e^2 + 4 \left(F_+ + i F_{\times} \right),
\nonumber \\
\xi_{3} &=& 
\left(- \frac{39934951}{122880} F_+ -  \frac{40111999}{122880} i F_{\times} \right) e^7 +
 \left( \frac{368823}{2560} F_+ + \frac{370047}{2560} i F_{\times} \right)e^5 
\nonumber \\
&+& \frac{813}{16} \left(- F_+ - i F_{\times}\right) e^3 + 9
\left(F_+ + i F_{\times} \right) e,
\nonumber \\
\xi_{4} &=& \left(- \frac{6662759}{10752} F_+ - \frac{300571739}{483840} i F_{\times} \right) e^8 + 
  \left(\frac{389167}{1440} i F_{\times} + \frac{388463}{1440} F_+\right) e^6 
\nonumber \\
&+&
 \frac{277}{3} \left(-F_+ - i F_{\times} \right) e^4 + 16
\left(F_+ + i F_{\times} \right) e^2,
\nonumber \\
\xi_{5} &=&  \left(  \frac{13445625}{28672} F_+ 
+ i \frac{13460625}{28672} F_{\times} \right) e^7 - \frac{89375}{576}\left( i F_{\times} + F_+ \right) e^5 +
\frac{625}{24} \left(F_+ + i F_{\times}\right) e^3,
\nonumber \\
\nonumber \\
\xi_{6} &=& \left( \frac{5605821}{7168} F_+ + \frac{28049841}{35840}
  i F_{\times} \right) e^8 - \frac{39987}{160} \left( F_+ +
  i F_{\times} \right) e^6 + \frac{81}{2} \left( F_+ + i F_{\times}
\right) e^4,  
\nonumber \\
\xi_{7} &=& - \frac{117649}{92160} \left(-48 + 307 e^2\right) e^5 \left(F_+ + i
  F_{\times}\right), 
\nonumber \\
\xi_{8} &=& - \frac{256}{945} \left(-336 + 2227 e^2\right) e^6 \left(F_+ + i
  F_{\times}\right), 
\nonumber \\
\xi_{9} &=& \frac{4782969}{35840} e^7 \left(F_+ + i F_{\times}\right),
\nonumber \\
\xi_{10} &=& \frac{390625}{2016} e^8 \left(F_+ + i F_{\times}\right).
\label{xis}
\eeq
\end{widetext}

\section{\label{app:ISCO}Ending frequency for eccentric binaries}

In this Appendix we discuss possible generalizations of the notion of an ISCO
to eccentric binaries. The idea is that eccentric binaries will transition
from inspiral to plunge at a frequency slightly different from the circular
ISCO frequency, and we may worry about the effect of this modified ISCO on the
upper cut-off frequency used in SNR calculations.

A possible way to modify the ISCO location is to use the Newtonian formula
in Eq.~\eqref{c0} with $a = p/(1 - e^{2})$ and $p = 6 + 2 e$, which
corresponds to the value of the separatrix between stable and unstable
(plunging) orbits. In this way we would find that the ISCO frequency is
\be\label{Newtonian}
F_{\ISCO}=\f{1}{2\pi M}\left(\f{1-e^2}{6+2e}\right)^{3/2}\,.
\ee
This guess cannot be valid for large eccentricities, when the pericenter
becomes small, since then the Newtonian relations break down. A more accurate
approximation of the ISCO frequency is to use the pericenter frequency
$\Omega_{p}^{2} = M/r_{p^{3}}$ at the separatrix pericenter $r_{p} = 6 + 2 e$,
leading to~\cite{Cutler:1994pb,Levin:2008yp,Komorowski:2009cg}
\be\label{EMRI}
F_{\ISCO}=\f{1}{2\pi M}\left(\f{1+e}{6+2e}\right)^{3/2}\,,
\ee
but this result is also not appropriate here because this eccentricity
corresponds to that associated with Schwarzschild geodesics, so it is not
equivalent to the Newtonian definition of eccentricity we have used (see
e.g.~\cite{JenaEccentric07} for a discussion).

One expects the residual eccentricity any binary could have by the time it
enters the strong field to be small.  The classic work of
Peters~\cite{Peters:1963ux} and Peters and Mathews~\cite{Peters:1964zz}
suggests that a binary with some moderate initial eccentricity will rapidly
circularize. Since $e/e_{0} \sim (f/f_{0})^{-19/18}$ to leading order [see
  e.g.~Eq.~$(2.34)$ in~\cite{Mora:2003wt}], an orbit with initial eccentricity
$e_{0} = 0.4$ at the beginning of the LIGO band will have a final eccentricity
of $e\simeq 0.035$ by the time it reaches LIGO's highest sensitivity region at
$200$~Hz. These results suggest that the ISCO frequency for eccentric
inspirals will generically be close to the ISCO frequency for circular
inspirals, provided this frequency is much larger than the initial frequency
associated with the initial eccentricity. If the latter is not the case
(e.g.~if a binary with $e_{0} = 0.4$ at $F_{0} = 20$~Hz merges at $40$~Hz),
then one might have to worry about the precise definition of the ISCO, but in
such cases the SNR will be dominated by the merger waveform and not the
inspiral. We are thus justified to ignore eccentric corrections to the ISCO
and employ the usual circular-orbit ISCO expression in our SNR calculations.

\section{\label{Fyr-app}Frequency at a given time before merger}

LISA sources can easily orbit for more than one year in the LISA band. The
LISA mission, however, is not expected to last for more than a few years in
orbit. For this reason, it is customary to perform LISA SNR and parameter
estimation calculations assuming that the source is observed over the last
year (or few years) of inspiral.

In the case of circular inspirals, one can compute exactly (to Newtonian
order) the frequency at a given time $T$ prior to merger. This is given by
Eq.~$(2.15)$ in Ref.~\cite{BBW05a}:
\be
f_{\rm yr} = 4.149 \times 10^{-5} \left(\frac{{\cal{M}}}{10^6 \; M_{\odot}} \right)^{-5/8} 
\left(\frac{T}{1 \; {\rm yr}}\right)^{-3/8} {\rm{Hz}}.
\ee
This equation can be obtained by finding $T(F)$ as the integral of
$\dot{F}^{-1}$ and then inverting the resulting expression to find $F(T)$.

In the case of eccentric inspirals, an analogous relation cannot be obtained
analytically.  This is because the equation for $\dot{F}$ in
Eq.~\eqref{fdotecc} is a function of the eccentricity, which itself is a
function of the frequency (and implicitly time). One could attempt to
construct an approximation for $F(T)$ by inserting Eq.~\eqref{eccentr} for
$e(F)$ into the expression for $\dot{F}$ in Eq.~\eqref{fdotecc} to compute
$T(F)$, and then perturbatively inverting this relation to find $F(T)$. The
resulting asymptotic series, however, is poorly convergent for large masses or
large integration times.

\begin{figure}[htb]
\includegraphics[width=8cm,clip=true]{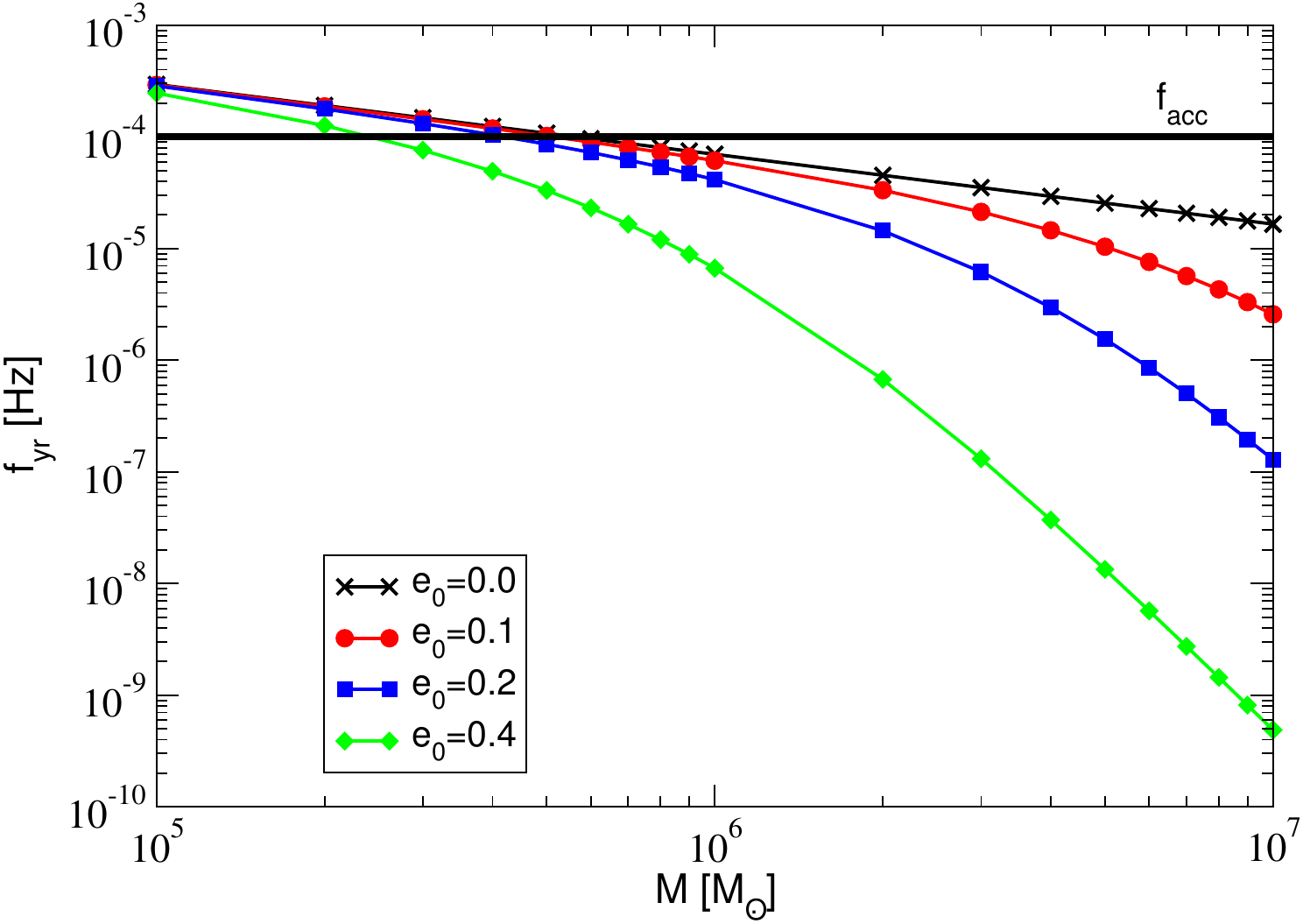}
\caption{\label{Fyr-plot} Frequency one year prior to merger as a function of
  total mass for equal mass binaries with different initial eccentricity:
  $e_{0} = 0$ (crosses), $e_{0} = 0.1$ (circles), $e_{0} = 0.2$ (squares)
  $e_{0} = 0.4$ (diamonds). The quantity $f_{\rm} = 10^{-4}$~Hz corresponds to
  the acceleration noise cut-off frequency.}
\end{figure}

A numerical procedure is thus necessary to find $f_{\rm yr}$ for eccentric
inspirals. One such scheme is as follows. Given some $e_{0}$ and $F_{0}$, one
can find the corresponding initial semi-major axis $a_{0}$ from the first
equality in Eq.~\eqref{c0}. From this, one can then use this same equation to
find the corresponding constant $c_{0}$. The eccentricity $e_{T}$ a time $T$
before merger is then given by Eq.~$(5.14)$ in~\cite{Peters:1963ux}, namely
\be
T(a_{0},e_{0}) = \frac{12}{19} \frac{c_{0}^{4}}{\beta} \int_{0}^{e_{T}} \frac{de \; e^{29/19}}{\left(1 - e^{2}\right)^{3/2}} \left(1 + \frac{121}{304} e^{2} \right)^{1181/2299},
\label{Tyr}
\ee
where $\beta = m_{1} m_{2} M$. This is because the eccentricity at zero
orbital separation (roughly corresponding to ``merger'') vanishes in the
Newtonian approximation.  If we set $T = 1 \; {\textrm{yr}}$, then we can
solve Eq.~\eqref{Tyr} for $e_{\textrm{yr}}$ numerically using bisection or the
secant method in Mathematica. Once the appropriate $e_{\textrm{yr}}$ is found,
one can use this to find $a_{\rm yr}$ via Eq.~\eqref{c0}, which can then also
be used to find $f_{\rm yr}$ for some given $e_{0}$ and $M$.

Figure~\ref{Fyr-plot} plots the dominant GW frequency (twice the orbital
frequency) one year prior to merger found with the above algorithm as a
function of total mass for equal mass binaries with different initial
eccentricities.  Observe that as the eccentricity increases, $f_{\rm yr}$
decreases faster with total mass than in the circular case, because
eccentricity speeds up the inspiral. Thus, given a fixed inspiral time
(e.g.~one year), the starting frequency must be pushed to lower values.  Care
must be taken, however, since $f_{\rm yr}$ appears multiplied by $\ell/2$ in
the step functions used to truncate the waveform. Nonetheless, the above
figure suggests that for total masses $M \gtrsim 10^{6} M_{\odot}$ it does not
matter whether one uses the circular or eccentric expressions for $f_{\rm
  yr}$.


\bibliography{ref-list}
\end{document}